\documentclass[11pt,letterpaper]{article}
\usepackage[T1]{fontenc}
\usepackage[utf8]{inputenc}
\usepackage{lmodern}
\usepackage[letterpaper,margin=1in,headheight=14pt]{geometry}
\usepackage{amsmath,amssymb,amsthm,mathtools}
\usepackage{microtype}
\usepackage{enumitem}
\usepackage{booktabs,array,tabularx,needspace,tikz}
\usetikzlibrary{arrows.meta,positioning,shapes.geometric}
\usepackage[hidelinks]{hyperref}
\hypersetup{pdftitle={Greedy BST is O(log log n)-Competitive},pdfsubject={Full conference submission},pdfkeywords={binary search trees, Greedy, competitive analysis, dynamic optimality},pdfauthor={Moshe Lewenstein}}
\usepackage{aliascnt}
\usepackage[nameinlink,noabbrev]{cleveref}
\usepackage{fancyhdr}
\setlist[enumerate]{itemsep=0.2em,topsep=0.4em}
\setlist[itemize]{itemsep=0.2em,topsep=0.4em}
\numberwithin{equation}{section}
\newtheorem{theorem}{Theorem}[section]
\newaliascnt{lemma}{theorem}
\newtheorem{lemma}[lemma]{Lemma}
\aliascntresetthe{lemma}
\crefname{lemma}{Lemma}{Lemmas}
\Crefname{lemma}{Lemma}{Lemmas}
\newaliascnt{proposition}{theorem}
\newtheorem{proposition}[proposition]{Proposition}
\aliascntresetthe{proposition}
\crefname{proposition}{Proposition}{Propositions}
\Crefname{proposition}{Proposition}{Propositions}
\newaliascnt{corollary}{theorem}
\newtheorem{corollary}[corollary]{Corollary}
\aliascntresetthe{corollary}
\crefname{corollary}{Corollary}{Corollaries}
\Crefname{corollary}{Corollary}{Corollaries}
\crefname{theorem}{Theorem}{Theorems}
\Crefname{theorem}{Theorem}{Theorems}
\crefname{section}{Section}{Sections}
\Crefname{section}{Section}{Sections}
\crefname{figure}{Figure}{Figures}
\Crefname{figure}{Figure}{Figures}
\crefname{table}{Table}{Tables}
\Crefname{table}{Table}{Tables}
\theoremstyle{definition}
\newtheorem{definition}[theorem]{Definition}
\theoremstyle{remark}

\DeclareMathOperator{\OPT}{OPT}
\DeclareMathOperator{\Greedy}{Greedy}
\DeclareMathOperator{\depth}{depth}
\DeclareMathOperator{\sgn}{sgn}
\DeclareMathOperator{\bitlen}{bitlen}
\DeclareMathOperator{\LCA}{LCA}
\newcommand{\Norm}{\mathcal N}
\DeclareMathOperator*{\argmin}{arg\,min}
\newcommand{\HH}{\mathcal H}
\newcommand{\EE}{\mathcal E}
\newcommand{\Act}{\mathcal A}
\newcommand{\VV}{\mathcal V}
\newcommand{\Neq}{N_{\mathrm{eq}}}
\newcommand{\Tint}{T_{\mathrm{int}}}

\newcommand{\ind}[1]{\mathbf 1_{\{#1\}}}
\newcommand{\llog}{\lambda}
\title{\LARGE Greedy BST is $O(\log\log n)$-Competitive}
\author{Moshe Lewenstein\thanks{Work on this paper was supported by ISF grant \#~2760/25 awarded by the Israel Science Foundation.}\\[0.4em]\normalsize Department of Computer Science\\\normalsize Bar-Ilan University, Israel}
\date{}

\hypersetup{bookmarksdepth=3}
\makeatletter
\renewcommand*\l@subsection{\@dottedtocline{2}{1.5em}{2.8em}}
\renewcommand*\l@subsubsection{\@dottedtocline{3}{4.3em}{3.6em}}
\makeatother
\tikzset{keynode/.style={circle,draw,inner sep=0pt,minimum size=5.5mm,font=\normalsize},
 ed/.style={line width=0.55pt},shorted/.style={ed,dashed},longed/.style={ed,line width=1.05pt}}

\begin{document}
\begin{titlepage}
\thispagestyle{empty}
\renewcommand{\footnotesize}{\normalsize}
\maketitle
\thispagestyle{empty}
\renewenvironment{abstract}{\par\vspace{1em}\begin{center}\bfseries Abstract\end{center}\normalsize}{\par}
\begin{abstract}

Greedy is a principal candidate for dynamically optimal binary search
trees. We prove that online Greedy is \(O(\log\log n)\)-competitive,
improving the \(2^{O(\sqrt{\log\log n})}\) bound of Guo, Pettie, Skora, and Wan~\cite{GPSW26}. For every
access sequence \(X\) on \(n\) keys and every initial tree \(T_0\), we
obtain

\[\operatorname{Greedy}(X,T_0)=O\!\left(n+\operatorname{OPT}_n(X)\log\log n\right).\]
Here \(\operatorname{OPT}_n(X)\) is the minimum cost of an offline BST
execution with a freely chosen initial tree.

We further show that the competitive factor depends only on the number
\(k\) of distinct requested keys:
\[
\operatorname{Greedy}(X,T_0)
=O\!\left(n+\operatorname{OPT}_n(X)\log\log k\right).
\]
The additive \(O(n)\) term accounts for an arbitrary initial tree.

\end{abstract}
\end{titlepage}
\pagenumbering{arabic}

\section{Introduction}\label{sec:intro}

Binary search trees are among the most basic data structures for
maintaining an ordered set. A balanced tree on \(n\) keys supports a
search in \(O(\log n)\) worst-case time, which is optimal for a single
search in the comparison model. Over a sequence of searches, however,
this guarantee can be far from optimal. If the same key is requested
repeatedly, keeping it at the root gives constant cost per request. More
generally, a tree that adapts to the access sequence can exploit
regularity that a worst-case guarantee does not capture. The dynamic
optimality problem asks how well an online BST can perform relative to
an optimal offline BST that knows the entire sequence in advance and may
change its shape as the sequence is served.

This problem originated with the splay trees of Sleator and Tarjan
\cite{ST85}. Introduced in 1985, splay trees use fixed rotation rules to
move each accessed key to the root. Sleator and Tarjan conjectured that
this simple self-adjusting strategy is dynamically optimal: its total
cost is within a constant factor of the offline optimum, apart from an
allowance for initialization. A second principal candidate is Greedy,
introduced by Lucas \cite{Luc88} in 1988 and independently rediscovered by
Munro \cite{Mun00} in 2000. Its original formulation is offline: after an
access, it rearranges the search path to favor subsequent requests.
\cite{DHIKP09} gave a geometric characterization of BST
executions and an online realization of the corresponding Greedy process
with only constant-factor overhead. We study this online geometric form
of Greedy. For the original offline formulation, GreedyFuture, Sadeh and
Kaplan~\cite{SK23} proved a lower bound of $2$ on its competitive ratio
and showed that its additive gap from $\OPT$ can be
$\Omega(m\log\log n)$ on sequences of $m$ requests to $n$ keys.

Progress toward dynamic optimality has followed two related directions.
One designs BST algorithms specifically to obtain a provable competitive
ratio. Tango trees \cite{DHIP07} achieve an \(O(\log\log n)\)
factor, as do the subsequent multi-splay trees \cite{WDS06} and chain-splay
trees \cite{Geo08}. Their analyses compare the algorithm's cost with
Wilber's interleave lower bound \cite{Wil89}. Lecomte and
Weinstein~\cite{LW20} showed that this bound, also called the alternation
bound, can underestimate $\OPT$ by a factor of $\Omega(\log\log n)$ on
some access sequences.
The other direction analyzes
splay and Greedy themselves. Both algorithms are known to exploit
several forms of temporal and spatial locality and to perform well on
structured access sequences. In particular, working-set bounds capture
temporal locality through logarithmic dependence on one plus the number
of distinct keys requested since the previous access to the same key
\cite{ST85,Fox11}, while dynamic-finger bounds capture spatial locality
through logarithmic dependence on one plus the rank distance between
consecutive requests \cite{CMSS00,Cole00,IL16}; these are amortized guarantees
with an allowance for initialization. Both algorithms also access all
\(n\) keys in sorted order in \(O(n)\) time \cite{Fox11,Tar85} and have
near-linear bounds for deque sequences \cite{Pet08,CGJOPY23}. For a more
complete picture of the history, see~\cite{Iac13}, \cite{GPSW26}, and~\cite{CHH26}.
Such guarantees support the dynamic optimality conjectures, but do not by themselves
establish a sublogarithmic competitive ratio on arbitrary sequences.
Until the recent developments described below, neither candidate was
known to be \(o(\log n)\)-competitive.

\subsection{Related work and recent progress}\label{sec:recent}

The first sublogarithmic bound for splay was proved in~\cite{CHH26}.
The \(O(\log\log n\, (\log\log\log n)^2)\) guarantee follows from a
direct comparison with a \emph{comparator}, an evolving near-optimal tree. Subsequently, in
independent work, Guo, Pettie, Skora, and Wan \cite{GPSW26} established the
first sublogarithmic competitive ratio for Greedy, proving a
\(2^{O(\sqrt{\log\log n})}\) bound. Their analysis relates geometric
cost measures at finer and coarser scales and ultimately charges the
cost to Wilber's interleave bound.

These results establish nontrivial competitiveness for both principal
candidates. We obtain the stronger \(O(\log\log n)\) bound for Greedy.
The proof develops a direct structural and amortized
analysis of Greedy's accesses against a moving comparator, using the
normal form of~\cite{CHH26}.

\subsection{Main results}\label{sec:result}

Our main theorem bounds Greedy against the offline optimum for every
access sequence and every initial tree. Fix the ordered universe
\([n]=\{1,\ldots,n\}\) and an access sequence \(X\in[n]^m\). Let
\(\operatorname{OPT}_n(X)\) be the minimum cost of an offline BST
execution serving \(X\), with its initial tree chosen freely.

We measure Greedy's work using the geometric representation of BST
executions~\cite{DHIKP09}. A point \((v,t)\) records that key \(v\) is touched while
serving the \(t\)th request: horizontal position is key order and vertical
position is time. Each request contributes its own point, and every
additional key touched contributes another point on the same row. We
call each such point a \emph{geometric touch}. Let
\(\operatorname{Greedy}(X,T_0)\) denote the total number of these touches
when Greedy starts from \(T_0\); points encoding the initial tree are
not charged. The online geometric/BST correspondence
turns this touch bound into a BST cost bound with constant-factor
overhead. \Cref{sec:bstmodel} gives the precise geometric rule and
correspondence.

All logarithms are binary, and we write
\(\lambda(s)=1+\log_2\log_2(s+2)\) to include small universes.
In asymptotic bounds we write \(\log\log s\) in place of
\(\lambda(s)\), with a constant factor understood when \(s\le2\).
For real $z$, write $[z]_+=\max\{z,0\}$.

\begin{theorem}\label{thm:main}
There is an absolute constant \(c\) such
that, for every \(n\geq1\), every initial BST \(T_0\) on \([n]\), and
every access sequence \(X\in[n]^m\),
\begin{equation}\label{eq:main}
\operatorname{Greedy}(X,T_0)\leq c\bigl(n+\operatorname{OPT}_n(X)\lambda(n)\bigr).
\end{equation}
\end{theorem}

For \(m\geq n\), the bound \(\operatorname{OPT}_n(X)\geq m\) absorbs the
initialization term, giving \(O(\log\log n)\)-competitiveness. The
constant-factor online BST realization of geometric Greedy has the same
asymptotic guarantee. The additive \(O(n)\) term accommodates an
arbitrary initial tree: the first request can touch a path of length
\(n\), whereas a freely initialized comparator can place that key at its
root.

The competitive factor can be made to depend only on the number of
distinct requested keys, even when the tree stores many other keys.

\begin{theorem}[Active-alphabet bound]\label{thm:mainalphabet}
There is an absolute constant \(c'\) such that, for every \(n\geq1\),
every initial BST \(T_0\) on \([n]\), and every nonempty access sequence
\(X\in[n]^m\) containing \(k\) distinct requested keys,
\begin{equation}\label{eq:mainalphabet}
\operatorname{Greedy}(X,T_0)
\leq c'\bigl(n+\operatorname{OPT}_n(X)\lambda(k)\bigr).
\end{equation}
\end{theorem}

\Cref{thm:alphabet} proves a more precise version of
\cref{thm:mainalphabet}, with an initialization term determined by the
initial visibility profile. That quantity and the proof are developed in
\cref{sec:quotient}.

\Cref{thm:main} improves the competitive guarantee for Greedy
from \(2^{O(\sqrt{\log\log n})}\)~\cite{GPSW26} to \(O(\log\log n)\).
This matches the best known competitive guarantee for binary search
trees and improves on the previously known bounds for Greedy and splay
trees, the two principal candidates for dynamic optimality.

\Needspace{14\baselineskip}
\subsection{Contributions}\label{sec:overview}
Our proof builds on the comparison framework developed
in~\cite{CHH26}. We use its comparator normal form as a black box
(\cref{thm:normal}), and draw on its block organization, normalized
point gaps, logarithmic labels, and parts of the historical accounting
with creator records. Applying this framework to Greedy poses new
structural and amortized challenges. We resolve them through several
new results: an exact representation of Greedy's geometric accesses,
bounds on repeated block participation and historical charges, and a
capped-record potential. These ingredients are central to obtaining
the single $O(\log\log n)$ factor.

Our main result is a direct $O(\log\log n)$ competitive analysis of
online Greedy. The central difficulty is to charge its geometric
touches to the work of a separately evolving near-optimal comparator:
one access may touch many keys, and one comparator rotation may change
many depths. We develop an exact representation compatible with both
evolutions and prove the required amortized bounds. The comparator is
used only in the analysis; Greedy needs no knowledge of future requests.

Our first contribution is the structural connection between the two
executions. We represent each Greedy access by an ordered forest of
links in an auxiliary heap built from touch heights and comparator
ranks. This forest describes the resulting heap exactly and pays for
all geometric touches, up to a constant per request. We prove that
maintaining this representation across a comparator rotation at depth
$d$ creates $O(1+d)$ new block instances, even when many keys change
heap parents. Each access leaves at most two source blocks nonempty.
Counting the nonempty shrinking blocks before and after each event
therefore bounds total source-block participation by the number of
requests and the weighted comparator rotation cost.

Our second contribution is an amortized analysis that incurs only one
$O(\log\log n)$ factor. A link need not decrease a label,
and an old creator record can enter many later charges. We resolve
these difficulties using the ordered structure of Greedy's complete
access forest. With one frozen snapshot per access, the source-block
participation bound and ordered cuts control both total label variation
and the sum of label differences across block boundaries. Bounding those differences
together avoids paying the maximum label separately for each boundary
link, which would introduce a second logarithmic factor. Bounded
historical reuse and a capped-record potential then account for the
remaining internal links.

Our third contribution separates initialization from the cost of
serving the requested keys. We prove that replacing each maximal interval
containing no requested key by one maximum-height representative gives
a geometric execution that tracks exactly which intervals Greedy
touches. This exact quotient, together with the visibility-loss bound,
pays for extra touches inside inactive intervals. The resulting bound
reduces the initialization term from
$O(n\log\log n)$ to $O(n)$ and makes the competitive factor
$O(\log\log k)$, where $k$ is the number of distinct requested keys:
$\Greedy(X,T_0)=O(n+\OPT_n(X)\log\log k)$.
\Cref{sec:strategy} presents the proof overview before the structural
and amortized arguments.

\Needspace{5\baselineskip}
\paragraph{Paper organization.}\label{sec:roadmap}
After the preliminaries in \cref{sec:model}, \cref{sec:strategy}
gives an overview of the analysis. \Cref{sec:structure,sec:amortized} prove the
structural and amortized bounds; \cref{sec:quotient} proves the
initialization and active-alphabet refinement. \Cref{sec:conclusion}
discusses the results' scope. All proofs of our new results are in the main text. The four-key example in
\cref{sec:workedgreedy} recurs throughout the proof; the larger example
in \cref{sec:workedcut} illustrates cut cancellation and boundary estimates.

\section{Preliminaries}\label{sec:model}
We first define the BST cost model and the geometric description of
Greedy. Following~\cite{DHIKP09}, the geometry records
\emph{which keys are touched at each time}, omitting intermediate
rotations. We then introduce the offline comparator and the parameters
used to bound Greedy's cost in terms of the comparator's work.

\subsection{BST model and geometric execution}\label{sec:bstmodel}
\subsubsection{Cost model}
We use the BST model defined in~\cite{DHIKP09}.
A BST stores one key at each node, with the keys in inorder. We consider
the searches on the fixed universe $[n]$. A search begins at
the root and must visit the requested key. We charge one unit for each
distinct node touched during the access, including the requested node;
repeated visits to the same node during that access are counted only
once. If $S_t$ is the set of keys touched while serving request $x_t$,
the access costs $|S_t|$, and the total cost is $\sum_{t=1}^m |S_t|$.

To specify the allowed rearrangements, the touched nodes $S_t$ must form
a connected subtree of the current BST containing both the root and
$x_t$. The algorithm may rearrange these nodes into any BST on the same
keys. Subtrees outside $S_t$ retain their internal structure and are
reattached in the positions required by key order. For example, touching
the three nodes of a root-to-key path and rearranging that path costs
three, regardless of the final shape. This formulation is
constant-factor equivalent to the usual model that charges individual
pointer moves and rotations, since a rearrangement of $|S_t|$ nodes can
be implemented with $O(|S_t|)$ such operations.

Keys are treated as atomic, with constant-time comparisons. Extensions
of comparison-driven structures to unbounded-length keys, accounting
for the cost of string comparisons, are studied in~\cite{AFGKLL14}.

\subsubsection{Geometric representation}
We first describe how an already specified BST execution is represented
geometrically, following~\cite{DHIKP09}. If the execution has
touch sets $S_1,\ldots,S_m$, its representation is the point set
\[
 P=\{(v,t):1\leq t\leq m,\ v\in S_t\}.
\]
The horizontal coordinate identifies the key, and the vertical coordinate
identifies the request time, with later times higher in the diagram.
Row $t$ therefore records exactly the keys touched during the $t$th
access. It contains the request point $(x_t,t)$ and one point for each
other touched key. Thus $|P|=\sum_{t=1}^m|S_t|$ is exactly the
node-touch cost. The representation omits the order of visits and the
particular rearrangement of the touched subtree. We distinguish
\emph{accessed}, meaning requested, from \emph{touched}, which includes
every key visited during that access.

Not every point set containing the request points represents a valid
BST execution. The following rectangle condition characterizes the
valid sets when the initial tree may be chosen freely.

\Needspace{4\baselineskip}
\begin{definition}[Arboreal satisfaction]
A finite point set is \emph{arboreally satisfied} if every pair on different rows and columns has a third point in its closed axis-aligned rectangle. Pairs on one row or one column are already satisfied. A point on the rectangle's boundary counts as a witness.
\end{definition}

For example, the two-point set $\{(1,1),(3,2)\}$ is unsatisfied, whereas
$\{(1,1),(3,2),(1,2)\}$ is satisfied. The point $(1,2)$ witnesses the
diagonal pair on the top edge of its rectangle. This boundary convention
is important later when equal heights block visibility.

More precisely, \cite{DHIKP09} show that a BST
execution yields an arboreally satisfied point set, and that every such
set containing the requests can be realized by a BST execution with the
same touch cost and a suitable initial tree. Thus the geometric view
preserves the cost while leaving the intermediate tree shapes
unspecified. This correspondence will also let us construct an offline
comparator from a satisfied point set.

\subsubsection{Online geometric Greedy}\label{sec:geometricgreedy}
The online geometric Greedy algorithm constructs the touch sets one
request at a time, using the rule of~\cite{DHIKP09}.
To specify its initial state, include one point
$(v,-\depth_{T_0}(v))$ for each key $v$, as
in~\cite{GPSW26}. These nonpositive-height points encode
$T_0$ and are not charged to the execution.

A horizontal sweep line moves upwards, processing requests in increasing
time order $t=1,2,\ldots,m$. On receiving $x_t$, include its request
point $(x_t,t)$ and examine the points below the current row. Select
every key $v\ne x_t$ for which an earlier point $(v,s)$ forms a closed
rectangle with $(x_t,t)$ containing no point other than these two
corners. All tests use the same history, before any extra points are
inserted on row $t$. Once selection is complete, add $(v,t)$ for every
selected key $v$. Together with the requested key $x_t$, these are the
keys touched during the access.

\Cref{sec:visibility} gives an explicit scan order for finding this set
using only the highest point in each column. Selection precedes
insertion throughout that procedure, so a newly inserted point cannot
suppress another required touch. Each access uses only the current
request and earlier touches. We measure Greedy's cost by its number of
geometric touches. The online equivalence theorem
of~\cite{DHIKP09} realizes this geometric algorithm as an
online BST with constant-factor cost overhead and a possible $O(n)$
setup charge. This setup charge is absorbed by the linear initialization
term in \cref{thm:main,thm:mainalphabet}.

\subsection{Height profiles and visibility}\label{sec:visibility}
The geometric history can be summarized by one height per key. Just
before a request, let $h(v)$ be the height of the highest point in
column $v$: its most recent touch time, or its initial height
$-\depth_{T_0}(v)$ if it has not yet been touched. An older point in
that column cannot define an empty rectangle with the new request,
because the highest point would already lie in the rectangle.

For a request to $x$ at time $t$, Greedy therefore touches $x$ and
every $v\ne x$ such that
\begin{equation}\label{eq:visible}
 h(z)<h(v)\quad\text{for every }z\in[\min\{x,v\},\max\{x,v\}]\setminus\{v\}.
\end{equation}
Thus the highest point in column $v$ must be strictly higher than the
highest points in all columns between $x$ and $v$, including column $x$.
A point at the same or a greater height in one of those columns lies in
the closed rectangle and prevents this extra touch. Equality blocks
because points on the boundary count.

The following conceptual scan identifies the touch set in
\eqref{eq:visible}. Select $x$, then scan left in the order
$x-1,x-2,\ldots,1$, maintaining a running maximum $b$ initially equal
to $h(x)$. At each key $v$, select it if $h(v)>b$, then set
$b=\max\{b,h(v)\}$. Scan right in the order $x+1,x+2,\ldots,n$,
restarting $b$ at $h(x)$. A selected key has a height larger than
every height encountered before it on that side, including $h(x)$;
we call such a height a \emph{strict outward height record}.

Keep all heights $h(v)$ unchanged during both scans. After the complete
touch set $S$ has been selected, set $h(v)=t$ for every $v\in S$ and
leave the heights of other keys unchanged. This update records the new
points $(v,t)$; it does not move any earlier point. Since $t$ is later
than every previous touch and every initial-state point, the selected
keys now all have the same height, larger than all remaining heights.

\subsubsection{Four-key execution}\label{sec:workedgreedy}
Let $T_0$ have root $1$, right child $2$, then right child $4$, with
$3$ as the left child of $4$. Its depths are $(0,1,3,2)$ in key order,
so the initial heights are $(0,-1,-3,-2)$.

\begin{figure}[!htbp]
\centering
\begin{minipage}[c]{0.30\textwidth}
\centering
\begin{tikzpicture}[x=0.70cm,y=0.73cm]
\node[keynode] (a) at (0,3) {$1$};
\node[keynode] (b) at (1,2) {$2$};
\node[keynode] (d) at (2,1) {$4$};
\node[keynode] (c) at (1,0) {$3$};
\draw[ed] (a)--(b)--(d)--(c);
\node[font=\normalsize,align=center] at (1,-0.8) {Initial BST $T_0$};
\end{tikzpicture}
\end{minipage}\hfill
\begin{minipage}[c]{0.65\textwidth}
\centering
\begin{tikzpicture}[x=0.85cm,y=0.62cm]
\draw[->] (0.6,-3.65)--(4.8,-3.65) node[right,font=\normalsize] {key};
\draw[->] (0.45,-3.3)--(0.45,2.7) node[above,font=\normalsize] {time};
\foreach \x in {1,2,3,4}{\node[font=\normalsize] at (\x,-3.98) {$\x$};}
\foreach \y in {-3,-2,-1,0,1,2}{\node[font=\normalsize,left] at (0.43,\y) {$\y$};}
\foreach \y in {1,2}{\draw[dotted] (0.65,\y)--(4.45,\y);}
\draw[dashed] (2,-1) rectangle (4,1);
\foreach \x/\y in {1/0,2/-1,3/-3,4/-2}{\draw[fill=white] (\x,\y) circle (2.3pt);}
\foreach \x/\y in {1/1,2/1}{\fill (\x,\y) circle (2pt);}
\node at (4,1) {$\times$}; \node at (2,2) {$\times$};
\node[font=\normalsize,align=center] at (2.8,-4.6) {$\circ$ initial\quad $\times$ request\quad $\bullet$ extra touch};
\end{tikzpicture}
\end{minipage}
\caption{The initial BST and its geometric encoding for requests $(4,2)$. Before the first row is completed, the dashed rectangle from $(2,-1)$ to $(4,1)$ is unsatisfied; the extra touch at $(2,1)$ repairs it. The rectangle from $(1,0)$ to $(4,1)$ is also initially unsatisfied and causes the touch at $(1,1)$. All first-row decisions use old heights. The second request is served without consulting any future request.}
\label{fig:geometric}
\end{figure}
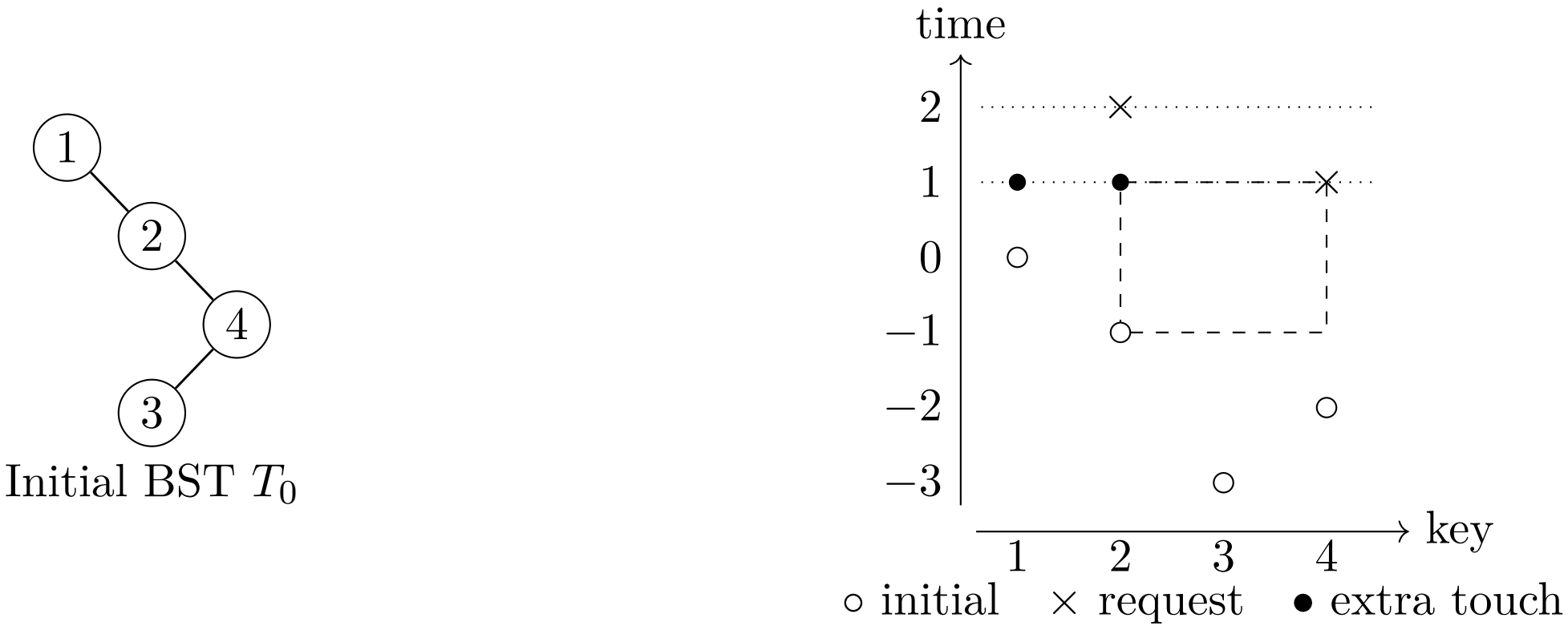

Request key $4$ at time $1$.
Select $4$ and start the left scan with threshold $b=h(4)=-2$.
Key $3$ is not selected because $h(3)=-3$ does not exceed $-2$.
Key $2$ is selected because $h(2)=-1>-2$, and the threshold becomes
$-1$. Key $1$ is then selected because $h(1)=0>-1$.
The touch set is therefore $\{1,2,4\}$.
Only now are those three heights set to $1$; $h(3)$ remains $-3$.

If the next request is $2$, each scan starts at threshold $h(2)=1$.
The other heights are $h(1)=1$, $h(3)=-3$, and $h(4)=1$; none is
strictly greater than $1$. Hence only $2$ is touched, and its height
becomes $2$.

\begin{center}
\begin{tabular}{@{}lrrrrl@{}}
\toprule
State & $h(1)$ & $h(2)$ & $h(3)$ & $h(4)$ & Touched in this access \\
\midrule
Initial & $0$ & $-1$ & $-3$ & $-2$ & --- \\
After requesting $4$ & $1$ & $1$ & $-3$ & $1$ & $\{1,2,4\}$ \\
After requesting $2$ & $1$ & $2$ & $-3$ & $1$ & $\{2\}$ \\
\bottomrule
\end{tabular}
\end{center}

\subsubsection{Height-order invariance}
The same selection rule is meaningful for any finite initial height
profile $h$, even if its values do not encode an initial BST. In this
generalized process, after each request set the selected heights to
one common value greater than every current height, leaving all other
heights unchanged. Write $C(X;h)$ for its total touch count. The next
lemma shows why the actual height values do not matter beyond their
relative order. This generality will be useful when we collapse
intervals of unrequested keys in \cref{sec:quotient}. For the intended
initialization,
\[
 \Greedy(X,T_0)=C(X;-\depth_{T_0}).
\]

\begin{lemma}[Height-order invariance]\label{lem:height}
Profiles with the same weak order, including ties, have the same touched sets. Raising their common touched set above all other heights preserves this equivalence.
\end{lemma}
\begin{proof}
The tests in \eqref{eq:visible} involve only comparisons of heights. After an access, the touched keys form a new maximum equivalence class; comparisons among untouched keys are unchanged.
\end{proof}

\subsection{Comparator normalization}\label{sec:comparator}
The comparator is a complete, separately evolving BST execution serving $X$, not a fixed reference tree. Its job is to provide ranks with which Greedy's work can be measured. We may choose it with knowledge of $X$ because it is used only in the proof.

The benchmark $\OPT_n(X)$ allows the offline optimum to choose its
initial BST freely, independently of Greedy's prescribed initial tree
$T_0$.

\paragraph{Comparator normal form.}
We use the following normal-form result of~\cite{CHH26}. At each
designated access state, the requested key is at the comparator's root.
\begin{theorem}[Normalized comparator]\label{thm:normal}
For an access sequence on $[n]$, there is a comparator execution whose designated access states have the requested keys at their roots, whose height is $O(\log(n+1))$ throughout, and whose rotations occur at depth $O(1)$. If it makes $R$ rotations and serves $m$ requests, then
\[
 R+m=O(\OPT_n(X)).
\]
\end{theorem}
Fix a normalized comparator execution supplied by \cref{thm:normal},
and let $T^*$ denote its current BST. For each request $x_t$, first
advance the comparator to its designated state with $x_t$ at the root.
Hold $T^*$ fixed while analyzing Greedy's response. Then allow the
comparator to move toward the next designated root. This serialization
is accounting, not communication between the executions. Greedy never
reads $T^*$ or its future plans.

We will bound Greedy's total touch cost in terms of the comparator's
requests and rotations, using the comparator depths to account for
changes in Greedy's state. Since the normalized comparator performs
$O(\OPT_n(X))$ work, this comparison yields a bound relative to the
offline optimum.

\subsection{Comparison parameters}\label{sec:parameters}
We build an auxiliary ordered heap $\HH(h,r)$ from Greedy's touch
heights $h$ and the comparator depths $r$, with the formal definition
given in \cref{sec:heap}. This heap is used only in the analysis.
The following table summarizes these objects and distinguishes
\emph{height} from \emph{rank}:
\begin{center}
\begin{tabularx}{\textwidth}{@{}p{0.20\textwidth}XX@{}}
\toprule
Object & Meaning & Changes when \\
\midrule
$h(v)$ & Most recent Greedy touch time of key $v$ & Greedy touches $v$ \\
$r(v)$ & Depth of $v$ in the comparator $T^*$ & The comparator rotates \\
$\HH(h,r)$ & Analysis-only ordered heap combining visibility and interval representatives & Either of the above changes \\
\bottomrule
\end{tabularx}
\end{center}
The auxiliary heap's parent relation is neither Greedy's physical parent relation nor, in general, the comparator's parent relation.

We state the comparison for any comparator $T^*$ having the requested
key at the root at each designated access time. This makes the
dependence on its height and rotation depths explicit; we then apply
the bound to the normalized comparator of \cref{thm:normal}. Two
quantities will control the proof: a budget $M$ for structural changes
and a bound $G$ on the number of logarithmic label levels.

Let $H$ be the maximum height of the comparator over the execution.
For the comparator's $i$th rotation, let $d_i$ be the depth in $T^*$ of
the key being promoted, measured \emph{immediately before that rotation}.
We weight this rotation by $1+d_i$ and define
\begin{equation}\label{eq:parameters}
 \EE=\sum_{i=1}^R(1+d_i),\qquad M=n+m+\EE.
\end{equation}
For the normalized comparator, each individual rotation has
$d_i=O(1)$, but bringing a requested key to the root can require
many such rotations. Every rotation contributes separately to $R$
and $\EE$, so $\EE=O(R)$ does not assert constant work per request.
We retain the weights $1+d_i$ for the more general comparison bound,
which also allows rotations at greater depths.
Thus $M$ includes $n$ units for initialization, $m$ for requests, and
the weighted rotation cost $\EE$. At each time the comparator rank is
$r(v)=\depth_{T^*}(v)$. In the analysis, a Greedy access changes the
heights $h$ while keeping $r$ fixed, and a comparator rotation changes
$r$ while keeping $h$ fixed.

The labels compress nonnegative gaps derived from comparator depths
into logarithmically many levels; the precise gaps are defined in
\cref{sec:blocks}. We use the same logarithmic transformation as the
contracted point gaps of~\cite{CHH26}. For a nonnegative
integer $q$, define
\[
 \bitlen(0)=0,\qquad \bitlen(q)=1+\lfloor\log_2q\rfloor\quad(q\ge1).
\]
The label function and its maximum over gaps bounded by $H$ are
\begin{equation}\label{eq:ranks}
 F(q)=1+\bitlen(q)=1+\lceil\log_2(1+q)\rceil,
 \qquad G=1+\bitlen(H).
\end{equation}
For example, gaps $0$, $1$, $2$--$3$, and $4$--$7$ receive labels
$1$, $2$, $3$, and $4$, respectively. The value $F(q)-1$ counts the
powers of two not exceeding $q$, so $G=O(1+\log(H+1))$.
For a normalized comparator, $M=O(n+\OPT_n(X))$ and
$G=O(\log\log n)$. The structural and amortized analysis will bound the
touch cost by $O(MG)$, as stated in the following theorem.

\begin{theorem}[Comparator-relative bound]\label{thm:comparison}
There is an absolute constant $c$ such that, for every finite initial
height profile $h$ and every comparator described above,
\begin{equation}\label{eq:comparison}
 C(X;h)\le c\,(n+m+\EE)G.
\end{equation}
\end{theorem}

The proof is completed in \cref{sec:closing}. Allowing arbitrary initial
heights is essential for \cref{sec:quotient}: after we collapse an
interval of unrequested keys to its maximum height, the resulting
profile may not encode a BST. The comparison bound still applies.

\section{Overview of the analysis}\label{sec:strategy}
The proof has two tasks: compare Greedy's work with the comparator's
movement, and then separate the initial tree's contribution. Recall
\[
 M=n+m+\EE,\qquad G=1+\bitlen(H),
\]
where $H$ is the comparator height and $\EE=\sum_i(1+d_i)$ is its
weighted rotation cost. For the normalized comparator,
$M=O(n+\OPT_n(X))$ and $G=O(\log\log n)$. We prove the base bound
$C(X;h)=O(MG)$ for every finite initial height profile $h$.

\subsection{Access forests and persistent blocks}\label{sec:strategygeometry}
The auxiliary heap combines Greedy's heights with comparator depths.
On an interval $J$, choose its shallowest comparator key $u$ as root.
Attach the keys visible from $u$ as singleton children; on each gap
between them, recursively construct the heap and attach its root as
another child. Order the children by key. Every subtree is an interval,
and an auxiliary parent is a comparator ancestor of its child.

Immediately before a request to $x$, the comparator has root $x$.
The auxiliary root is therefore also $x$, and its children represent
the touched singletons and the untouched gap intervals. Raising all
touched heights to a common maximum changes parents only among these
old children. Short links first attach some touched singletons to
neighboring gap representatives. Long links then form the minimum-depth
Cartesian tree on the remaining representatives on each side of $x$.
The old gaps retain all their internal edges. This is an exact
description of the new canonical heap, proved in \cref{sec:heap}.

The union of the short and long links is one forest on the old children,
with at most two roots. A link $v\to w$ makes $v$ the new parent of
$w$; we call $v$ its winner and $w$ its loser. Each key loses at most
once in the whole access, and each winner receives at most four new
children. If an access creates $e$ links and touches $K$, then
\[
 |K|\le e+3.
\]
Thus it suffices to count links. Short and long links are two geometric
parts of the same forest; the accounting will read one snapshot for both.

\label{sec:strategyblocks}
Consecutive same-side children are grouped into persistent blocks.
A growing block appends children in outward order. Once converted to
shrinking, it receives no further entries and never grows again.
Splits and transfers preserve the members' original entry records.
The current parent is the block's owner; its original owner is its
immutable creator. Each key has only one current membership.

A comparator rotation at depth $d$ needs $O(1+d)$ block creations,
including split fragments. The total number of block instances is
therefore $L=O(M)$. The rotation may move many keys, but the block
normalization described below absorbs uniform changes together.

Count a block--access incidence when a frozen source block participates
in a link between different source blocks. Let $Z$ be the number of
nonempty shrinking blocks. An access converts at most two growing
blocks and leaves at most two source blocks nonempty. If it has $N_t$
nonempty sources, then $N_t+\Delta_t Z\le4$, where $\Delta_t Z$
includes both conversion and assembly. Initially $Z=0$, and a
comparator rotation at depth $d$ increases $Z$ by at most $32(1+d)$.
Summing over events and using $Z\ge0$ gives the incidence budget
\[
 \Act\le4m+32\EE=O(M).
\]
This one population count supplies the structural budget used by all
boundary estimates. The \emph{complete flushing} property,
that each old source block has at most one member after assembly,
will also control reuse of historical records.

\subsection{Labels and boundary control}\label{sec:strategyboundary}
For a child $v$ of $u$, its raw gap is $r(v)-r(u)$. A shrinking block
subtracts its minimum raw gap from every member; the resulting values
are its point gaps. A growing block uses baseline zero. Uniform shifts
cancel under minimum subtraction, and restriction to a smaller block
can only decrease surviving point gaps. Our comparator maintenance uses
these facts to ensure that no primary label increases during a rotation.

The primary label is $g(v)=F(p(v))=1+\bitlen(p(v))\in[1,G]$.
Before an access, convert the root's growing blocks to shrinking and
update labels. Freeze those labels, source memberships, and records
for the complete short-and-long-link forest. A loser enters its winner's
growing block and records the winner's frozen label. Update primary
labels only after all entries of that access are complete.

A link is internal if its endpoints belong to the same frozen source
block; otherwise it is a boundary link. Let $I,B$ be their total counts,
and let $\Neq$ count equal-label internal links. An internal link with
old winner and loser point gaps $a<b$ gives its loser new gap $b-a$.
Its label cannot increase; if the endpoint labels are equal, it decreases.
Some other internal links make no such progress: gaps $0,1$ become
gap $1$, so the loser's label is unchanged.

Let $P^+,P^-$ be the total amounts of coordinatewise label increases
and decreases, and set $V=P^++P^-$. Only a boundary loser whose winner
lies below its source baseline in comparator rank can increase its
label. At a block endpoint, there is at most one such long link and at
most one short link in an access. Hence $P^+=O(G\Act)$.
Telescoping the sum of labels gives $V=O((n+\Act)G)=O(MG)$.

The same ordered-cut argument controls the remaining boundary work.
Long edges crossing a cut lie on one Cartesian-tree search path.
Their directions alternate, their ranks increase, and their rank
intervals have disjoint interiors. Alternation bounds signed label
sums; disjointness makes monotone label increments telescope. Short
links contribute at most one additional edge per block endpoint.
Writing
\[
 W=\sum_{v\to w\text{ boundary}}|g(v)-g(w)|,
\]
\cref{lem:boundarybudgets} proves
\[
 V=O(MG),\qquad B+\Neq=O(MG),\qquad W=O(MG).
\]
These estimates do not assume a bound on $I$. Bounding the weighted
quantity $W$ directly is essential: multiplying $B$ by $G$ would give
an unnecessary second logarithmic factor.

\subsection{Historical counting with capped records}\label{sec:strategyhistory}
To account for internal links without label decreases, each membership
stores the creator's frozen label at entry. Write this immutable number
as $c(w)$ for the current membership of $w$. Two geometric facts limit
reuse of the past. First, a membership wins at most four internal links,
all in its single nonsingleton flush. Second, a cut in a lineage's
entry times is a fixed cut in key order. At most one fragment straddles
it; its short and long internal crossings have signed total of magnitude
at most two, and flushing prevents another contribution later. Thus a
creator-label change has a bounded signed coefficient over the entire
history, rather than merely in one access.

The remaining record changes are handled by
\[
 \Gamma=\sum_{w\text{ currently in a block}}\min\{c(w),g(w)\}.
\]
This potential lies between zero and $(n-1)G$. On an internal link
$v\to w$, its record-replacement contribution is at least
$g(v)-c(w)$. On a boundary link, any negative contribution is at most
$[g(w)-g(v)]_+$. Comparator resets remove at most two records per
rotation, and primary-label changes reduce the potential by at most
$P^-$. Put $J_0=n-1+2R$, which bounds the number of initialization
and pivot entries. The signed internal record-change sum satisfies
\[
 \Tint:=\sum_{v\to w\text{ internal}}(g(v)-c(w))
 \le GJ_0+W+P^-.
\]
Capping the record at the current label allows this direct estimate even
when the historical record exceeds that label.

For an internal link, add and subtract the winner's entry label and
the two stored creator labels in an exact identity for its unit cost.
The resulting entry term is $1+c-a$, where $a$ is the winner's entry
label and $c$ its creator record. It is nonpositive for a membership
born at an unequal-label internal link, and is one for an equal-label
birth. Initial, pivot, and boundary births have direct bounds in terms
of $GJ_0$, $B$, and $W$. Applying the two reuse bounds above and the
capped-record estimate gives
\[
 I=O\!\left(GJ_0+B+\Neq+V+W\right).
\]
\Cref{sec:accounting} proves this bound with explicit constants.
Its right side is already
$O(MG)$, so $C(X;h)\le3m+I+B=O(MG)$.

\subsection{Initialization and active-alphabet reduction}\label{sec:strategyquotient}
The base bound is $O((n+\OPT_n(X))\log\log n)$. To separate
initialization, retain the $k$ requested keys and replace each maximal
inactive interval by one column at its maximum height. There are at most
$2k+1$ quotient columns, and the quotient Greedy execution tracks
exactly which intervals are touched.

An outside access to an inactive interval permanently removes all but
at most two of its touched keys from the union of its strict prefix
and suffix height records. Summing this visibility loss gives
\[
 C(X;h)\le(n-k)+2C(Y;\bar h),
\]
where $Y,\bar h$ are the quotient sequence and initial profile.
Monotone column projection of an optimal geometric execution preserves
arboreal satisfaction and does not increase its point count, so the
quotient optimum is $O(\OPT_n(X))$. The base comparison applies to
$\bar h$ even when that profile does not encode a BST. Since $k\le m$,
the quotient initialization term is absorbed by its optimum, giving
$O(n+\OPT_n(X)\log\log k)$. \Cref{sec:quotient} supplies the exact
statements and proofs.
\label{sec:overviewend}

\section{Structural analysis}\label{sec:structure}
\subsection{Auxiliary heap and access decomposition}\label{sec:heap}
Our first task is structural, not amortized: in the auxiliary heap, represent one Greedy access by an exact collection of simple parent changes. Visibility comes from the online execution; the representative of each untouched interval comes from the comparator. Counting these attachments will later pay for the touches.

\subsubsection{Cartesian trees and interval minima}
An interval needs a canonical representative that remains meaningful when the comparator changes. Choose its shallowest comparator key. A \emph{minimum Cartesian tree} is a BST ordered by key that is also heap-ordered by a prescribed priority: its minimum-priority key is the root, and the rule recurses on either side. It need not be balanced, and its priorities here are comparator depths, not random numbers.

Cartesian trees originate with Vuillemin~\cite{V80}; treap uniqueness is described by Seidel and Aragon~\cite{SA96}. We use minimum rather than maximum priorities. The interval-minimum/LCA correspondence is the usual RMQ--LCA equivalence~\cite{BF00}. For an exposition of the Cartesian-tree interpretation of range minima and its connection to lowest common ancestors, see~\cite{Lew13}.

\begin{lemma}[Cartesian uniqueness and interval minima]\label{lem:min}
Key order and a fixed priority order determine a unique Cartesian BST. A BST is the minimum Cartesian tree of its in-order depth sequence. In particular, for a nonempty key interval $[a,b]$ in a comparator $T^*$,
\[
 \argmin_{a\le z\le b}\depth_{T^*}(z)=\LCA_{T^*}(a,b).
\]
This minimizer is unique and is an ancestor of every key in the interval.
\end{lemma}
\begin{proof}
The minimum-priority key must be the root, and key order determines the
two recursive subproblems. Given a BST $T$, assign each key $v$ the
priority $\depth_T(v)$. Every parent in $T$ is strictly shallower than
its descendants, so this recursion recovers $T$ itself. In our
application, $T=T^*$, the current comparator. The search paths for $a$
and $b$ in $T^*$ coincide until reaching their lowest common ancestor.
This node lies in $[a,b]$ and is an ancestor of every key there, so it is
the interval's unique minimum-depth key.
\end{proof}
Incomparable keys may have equal depths. A consistent tie convention defines Cartesian trees for general priority sequences, but it changes neither the original comparator shape nor any numeric depth used in the analysis.

\Needspace{6\baselineskip}
\subsubsection{Canonical auxiliary heap}\label{sec:heapdefinition}
Visible keys alone would omit possible parents in the post-access structure. We therefore retain each nonempty invisible gap as a whole piece, represented by its minimum-rank key. A \emph{center} is the representative at which this recursive construction is currently rooted. Unlike a BST, the auxiliary heap can have many children on either side.

For a key $v$, define its height-blocking neighbors
\begin{align*}
 L_h(v)&=\max\bigl(\{u<v:h(u)\ge h(v)\}\cup\{0\}\bigr),\\
 R_h(v)&=\min\bigl(\{u>v:h(u)\ge h(v)\}\cup\{n+1\}\bigr),
\end{align*}
and its visibility neighborhood
\[
 N_h(v)=\{L_h(v)+1,\ldots,R_h(v)-1\}.
\]
By \eqref{eq:visible}, an access to $x$ touches $v$ exactly when $x\in N_h(v)$.

For a nonempty interval $J$, define $\HH(h,r;J)$ recursively. Choose its unique minimum-rank key $u$. Let $K(J,u)$ contain $u$ and the keys $v\in J\setminus\{u\}$ satisfying the outward-record condition \eqref{eq:visible} with $x=u$. In particular, each such $v$ has $h(v)>h(u)$. Make $u$ the root, make every key of $K(J,u)\setminus\{u\}$ a singleton child, and, for every nonempty gap between these keys, including exterior gaps in $J$, attach the root of the recursively constructed heap on that gap. Order the children by key. Write
\[
 \HH(h,r)=\HH(h,r;[n]).
\]
This is an analysis structure, not a physical representation maintained by Greedy. Its recursive roots are called \emph{centers}.

\begin{figure}[htbp]
\centering
\begin{minipage}[t]{0.30\textwidth}
\centering
\begin{tikzpicture}[x=0.78cm,y=0.65cm]
\path[use as bounding box] (-0.45,-0.4) rectangle (3.85,5.25);
\node[keynode] (c6) at (3.2,4.8) {$6$};
\node[keynode] (c2) at (1.2,3.2) {$2$};
\node[keynode] (c1) at (0,1.6) {$1$};
\node[keynode] (c4) at (2.2,1.6) {$4$};
\node[keynode] (c3) at (1.2,0) {$3$};
\node[keynode] (c5) at (3.2,0) {$5$};
\draw[ed] (c6)--(c2) (c2)--(c1) (c2)--(c4)
 (c4)--(c3) (c4)--(c5);
\end{tikzpicture}
\par\normalsize Comparator $T^*$
\end{minipage}\hfill
\begin{minipage}[t]{0.30\textwidth}
\centering
\begin{tikzpicture}[x=0.78cm,y=0.65cm]
\path[use as bounding box] (-0.45,-0.4) rectangle (3.85,5.25);
\node[keynode,fill=black!12] (b6) at (1.7,4.8) {$6$};
\node[keynode,fill=black!12] (b1) at (0,3.2) {$1$};
\node[keynode] (b2) at (1.7,3.2) {$2$};
\node[keynode,fill=black!12] (b5) at (3.4,3.2) {$5$};
\node[keynode] (b3) at (1.1,1.6) {$3$};
\node[keynode] (b4) at (2.3,1.6) {$4$};
\draw[ed] (b6)--(b1) (b6)--(b2) (b6)--(b5)
 (b2)--(b3) (b2)--(b4);
\node[font=\normalsize] at (1.7,0.4) {gap $\{2,3,4\}$};
\end{tikzpicture}
\par\normalsize Heap before request $6$
\end{minipage}\hfill
\begin{minipage}[t]{0.30\textwidth}
\centering
\begin{tikzpicture}[x=0.78cm,y=0.65cm]
\path[use as bounding box] (-0.45,-0.4) rectangle (3.85,5.25);
\node[keynode,fill=black!12] (a6) at (1.7,4.8) {$6$};
\node[keynode] (a2) at (1.7,3.2) {$2$};
\node[keynode,fill=black!12] (a1) at (0,1.6) {$1$};
\node[keynode] (a3) at (1.13,1.6) {$3$};
\node[keynode] (a4) at (2.27,1.6) {$4$};
\node[keynode,fill=black!12] (a5) at (3.4,1.6) {$5$};
\draw[ed] (a6)--(a2) (a2)--(a3) (a2)--(a4);
\draw[shorted] (a2)--(a1) (a2)--(a5);
\end{tikzpicture}
\par\normalsize Heap after request $6$
\end{minipage}
\caption{A recursive gap heap and its response to one access. Vertex
labels are keys. Shaded vertices in the heap panels are the keys
$\{1,5,6\}$ touched by request $6$. The middle panel retains the whole untouched
interval $\{2,3,4\}$ beneath its representative $2$; the right panel
adds the two dashed parent edges and preserves that interval's internal
edges. The comparator remains fixed throughout.}
\label{fig:heapconstruction}
\end{figure}
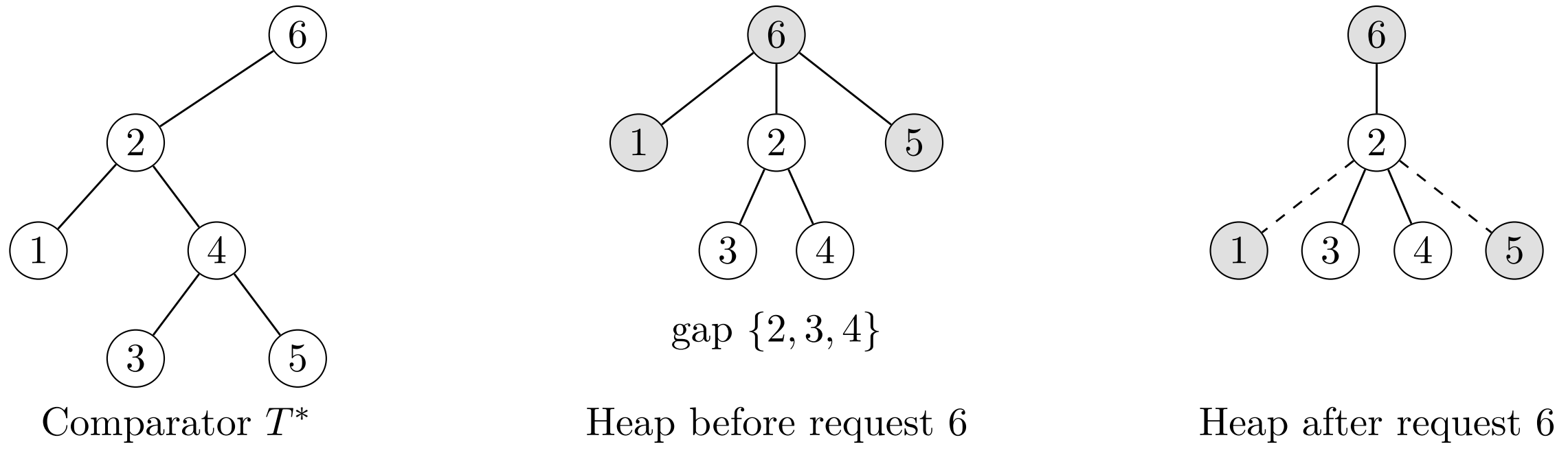

\paragraph{A six-key example.}
\Cref{fig:heapconstruction} shows the construction before and after a
request to $6$. Take the height and comparator-depth vectors, in key
order, to be
\[
 h=(0,-3,-2,-3,-1,-2),\qquad r=(2,1,3,2,3,0).
\]
The height profile can arise from an initial BST with root $1$, right
child $5$, children $3,6$ of $5$, and children $2,4$ of $3$, using
$h(v)=-\depth_{T_0}(v)$. The comparator is the different BST drawn in
the left panel.

Its shallowest key is $6$, so the auxiliary root is $6$. Scanning left
from $6$, the visible keys are $5$ and $1$: their heights $-1$ and $0$
are successive strict records above $h(6)=-2$. The intervening keys
$\{2,3,4\}$ form one gap. Its representative is $2$, whose comparator
depth $1$ is smallest in that interval. Within the recursive construction
on this gap, key $3$ is visible from $2$, while $\{4\}$ is a further
gap. Thus $6$ has children $1,2,5$, and $2$ has children $3,4$.
Here $3$ is a visible singleton in the recursive construction, although
it is not touched by the request to $6$.

That request touches $\{1,5,6\}$ and raises their heights to $1$.
All other keys now have height at most that of $6$, so none is visible
from $6$. Its sole gap is therefore $\{1,2,3,4,5\}$, represented by
$2$. Continuing the recursive construction gives the right panel:
$6$ has only child $2$, and $2$ has children $1,3,4,5$.
The old gap's edges $2\to3$ and $2\to4$ are preserved; only $1$ and
$5$, both formerly children of $6$, change parent. Their new parent
is $2$, as shown by the two dashed edges $2\to1$ and $2\to5$.
\Cref{sec:accessdecomposition} explains how to obtain this heap change
through two stages of links.

\begin{lemma}[Shielding and visibility centers]\label{lem:ownership}
Every key occurs once in $\HH(h,r)$. Every auxiliary subtree is a key interval. An auxiliary parent is a strict comparator ancestor of its child. Define the \emph{visibility center} of $v$ by
\[
 \operatorname{vc}(v)=\argmin_{z\in N_h(v)}r(z).
\]
Then $v$ is a center exactly when $\operatorname{vc}(v)=v$; otherwise it is a visible singleton child of $\operatorname{vc}(v)$.
\end{lemma}
\begin{proof}
The recursive construction partitions each interval into its center, visible singleton children, and gaps, so it covers every key once and gives interval subtrees. Its center is a comparator ancestor of every key in the interval by \cref{lem:min}. Every parent-child edge is consequently a strict comparator-ancestor edge.

Between consecutive outward records, the intervening heights are at most the inner record height, and the outer record is taller. In an exterior gap all heights are at most the last record height. Thus every finite bounding record has height at least every height in its adjacent gap. A neighborhood of a key inside that gap cannot cross the bounding record. Inducting through the recursion shows that its full neighborhood stays in its recursive interval.

For a visible child $v$ of a center $u$, visibility gives $u\in N_h(v)$, and the preceding shielding places $N_h(v)$ inside the same recursive interval. Hence $u$ is its minimum-rank key. For the center itself, its neighborhood contains it and stays in its recursive interval, so its minimum is itself. Recursion inside the gaps proves the characterization for every key.
\end{proof}
The visibility center is not always the auxiliary parent: a recursive
center has $\operatorname{vc}(v)=v$, even when it has a parent in the heap.

\subsubsection{Two-stage access decomposition}\label{sec:accessdecomposition}
We now make precise the short-link and long-link decomposition
introduced in \cref{sec:strategygeometry}.
We analyze an access to $x$ starting immediately after comparator
maintenance: the comparator has root $x$, and the auxiliary heap has
been updated to $\HH(h,r)$ for the pre-access Greedy heights $h$.
We hold the comparator ranks $r$ fixed throughout this access.
Neither of the two link stages below has yet been performed.

Let $K$ be the touched set, selected using $h$. The geometric access
raises all keys in $K$ to the same new maximum height, giving a new
profile $h'$. These are Greedy's latest-touch heights, used to define
the auxiliary heap, not depths in that heap. We will transform the
starting heap $\HH(h,r)$ into the post-access heap $\HH(h',r)$ by
short links followed by long links. In $h'$, equal-height blocking
allows a neighboring gap to see an immediately bordering touched key,
but not a more distant one beyond it. This explains the local short
links below. To prove that these two stages produce exactly the
canonical post-access heap $\HH(h',r)$, as asserted in
\cref{lem:assembly}, we must also show that every old untouched gap
retains all its internal auxiliary edges.

\Needspace{3\baselineskip}
Let $\VV$ be the ordered list of children of $x$ in the starting heap
$\HH(h,r)$. We call these the \emph{old children} throughout both link stages. On each side of $x$, the list partitions the keys on that side into \emph{atoms}. A T-atom is a touched-key singleton. A Q-atom is a whole old gap interval, represented by its minimum-rank key. Adjacent Q-atoms cannot occur, since consecutive gaps are separated by a touched key. Every consecutive union of atoms is a key interval with a unique minimum-rank representative.

For a T-atom $w$, let $A(w)$ be its immediate Q-atom neighbors in the original same-side atom list, and put
\[
 q(w)=\argmin_{z\in\{w\}\cup A(w)}r(z).
\]
Here a Q-atom is identified with its representative. The minimizer is unique by \cref{lem:min}.

\begin{lemma}[Local absorption]\label{lem:absorb}
After the access, a touched singleton $w\ne x$ is absorbed as a singleton child of a neighboring Q-representative precisely when $q(w)\ne w$. Its parent in that case is $q(w)$. All absorption decisions use the unchanged original list.
\end{lemma}
\begin{proof}
Give all touched keys their new common maximum height and consider canonical reconstruction on a segment of complete old atoms. If its minimum-rank representative is a T-atom, no other key is visible from it: other touched keys have equal height, and old-gap keys have lower height. Reconstruction chooses that T-atom and recurses on the left and right segments.

If the minimum is the representative $u$ of a Q-atom, its visible set inside its old interval is unchanged. The additional visible keys are exactly the bordering T-atoms that remain in the current segment. Each has the new maximum height and blocks all more distant touched keys. These bordering keys become singleton children of $u$. Internal gap subheaps are unchanged, and exterior recursion continues beyond the absorbed boundaries.

Fix $w$ and its zero, one, or two original Q-neighbors. Before one of this group is chosen as a reconstruction root, an outside root cannot separate the group, since no other atom lies between its members. Only an original neighboring Q-atom can absorb $w$. An outside root may remove a T-atom outside the group but cannot remove a Q-atom in the group. Thus all group members still belong to one segment when the first is selected. That member must be the group's minimum-rank representative, namely $q(w)$. If it is $w$, the key becomes a center; otherwise the selected Q-atom immediately absorbs $w$.
\end{proof}

In the two-stage access decomposition, each link $v\to w$ creates a
new auxiliary parent edge between two keys in the original child list
$\VV$, making $v$ the parent of $w$. We call $v$ the \emph{winner}
and $w$ the \emph{loser}. The short-link and long-link rules below
specify which links are created. These names describe a parent change,
not an actual comparison made by Greedy. Links are not asserted to
correspond one-for-one to physical BST rotations.

Call the absorption edges \emph{short links}. Contract each absorbed T-atom into its winning Q-atom, retaining the Q-representative. On the remaining representatives on each side, construct the minimum-rank Cartesian BST. Its parent-child edges are \emph{long links}, and its root remains a child of $x$.

\paragraph{Returning to the six-key example.}
In \cref{fig:heapconstruction}, the original atoms to the left of
request $6$ are the T-atom $\{1\}$, the Q-atom $\{2,3,4\}$ represented
by $2$, and the T-atom $\{5\}$. Since $r(2)=1<r(1)=2$ and
$r(2)=1<r(5)=3$, both absorption decisions select $2$. Thus the two
dashed edges $2\to1$ and $2\to5$ are short links. Both decisions use
the original atom list: the gap absorbs one touched singleton from
each side. The singletons lie on opposite sides of the gap, although
both are to the left of the requested key $6$.
After these absorptions, only representative $2$ remains on that side,
so the long-link stage adds no edges. The edge $6\to2$ is inherited
from the starting heap. \Cref{sec:workedassembly} gives a complementary
example using both short and long links.

\begin{lemma}[Exact two-stage assembly]\label{lem:assembly}
The short and long links have the following properties.
\begin{enumerate}[label=(\roman*)]
\item Every old Q-atom retains all its internal auxiliary edges. Only old children of $x$ receive new parents.
\item Each short link joins adjacent members of the original list $\VV$. Each representative receives at most one short child on each side.
\item After deleting the short losers from $\VV$, the long links form the two Cartesian BSTs just described.
\item A recipient receives at most one new child per side in each stage. New children are outside its old children, and a short child precedes a long child in outward order on the same side.
\item If $e$ is the number of short and long links, then $|K|\le e+3$.
\end{enumerate}
\end{lemma}
\begin{proof}
By \cref{lem:absorb}, each touched singleton has a unique absorption decision. Each contracted Q-piece consists of its old interval and at most its two original bordering singletons, so the pieces are disjoint consecutive intervals. Their representatives are still their minimum-rank keys. Remaining T-atoms are singleton pieces.

The Cartesian construction preserves the comparator-ancestor property.
Each recursive subproblem consists of consecutive pieces whose union is
a key interval $J$. Its minimum-rank representative is also the
minimum-rank key of $J$, so by \cref{lem:min} it is a comparator
ancestor of every key in $J$. Thus every long link $v\to w$ makes a
strict comparator ancestor $v$ the auxiliary parent of $w$. Short links
satisfy the same property because each contracted piece retains its
minimum-rank key.

The reconstruction in the preceding proof chooses the minimum representative, preserves every old Q-subheap, and recurses on the pieces to either side. Cartesian uniqueness (\cref{lem:min}) identifies this reconstructed skeleton; merely defining a candidate Cartesian tree would not suffice. Short children are at the immediate old boundaries of a Q-atom. A subsequent long child represents an exterior interval beyond those absorbed boundaries. This proves the first four assertions, including outward append order. A T-atom that was not absorbed has no old children and receives at most one long child on either side.

The short losers together with the residual representatives are exactly $\VV$. Restoring each short loser with its short parent edge gives a forest on $\VV$ with at most two roots. Hence $e\ge|\VV|-2$. Since $K\setminus\{x\}\subseteq\VV$, we obtain $|K|\le|\VV|+1\le e+3$.
\end{proof}

A long-link winner may also be a long-link loser; the analysis keeps the full forest.

\subsubsection{Four-key access decomposition}\label{sec:workedassembly}
Use the first access from \cref{sec:workedgreedy}, with $h=(0,-1,-3,-2)$ and request $x=4$. For this comparison choose $T^*$ with root $4$, left child $1$, $1$'s right child $3$, and $3$'s left child $2$. In key order its ranks are $r=(1,3,2,0)$.

The old auxiliary children of $4$ are $1,2,3$. Keys $1$ and $2$ are touched singletons; $3$ represents the untouched gap $\{3\}$. The only absorption is $3\to2$, because $3$ has smaller comparator rank than $2$. The residual representatives $1,3$ form the Cartesian tree with root $1$ and right child $3$, giving the long link $1\to3$. The unchanged request root $4$ retains just child $1$.

The visibility-center formula gives the same interpretation. Before the access,
$N_h(2)=\{2,3,4\}$ has minimum-rank key $4$, so $2$ is its visible
singleton child; $N_h(3)=\{3\}$ makes $3$ a center. Afterwards the
equal-height key $4$ blocks $2$, leaving $N_h(2)=\{2,3\}$ and new
visibility center $3$, which is also its new auxiliary parent. The
heights determine the neighborhood, while the comparator ranks select
its visibility center.

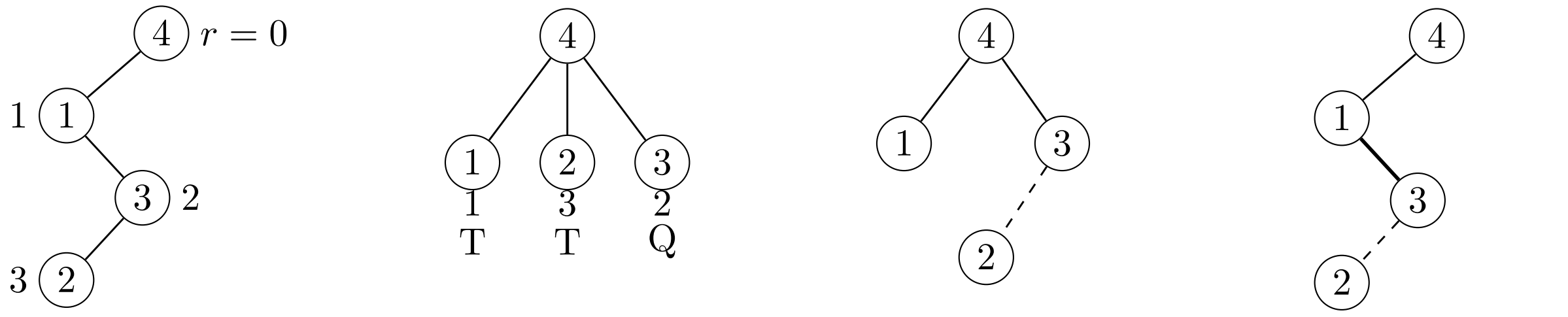
\begin{figure}[htbp]
\centering
\begin{minipage}[t]{0.23\textwidth}
\centering
\begin{tikzpicture}[x=0.64cm,y=0.64cm]
\node[keynode] (a4) at (2,3.9) {$4$};
\node[keynode] (a1) at (0.5,2.6) {$1$};
\node[keynode] (a3) at (1.7,1.3) {$3$};
\node[keynode] (a2) at (0.5,0) {$2$};
\draw[ed] (a4)--(a1)--(a3)--(a2);
\node[font=\normalsize,right] at (2.4,3.9) {$r=0$};
\node[font=\normalsize,left] at (0.1,2.6) {$1$};
\node[font=\normalsize,right] at (2.1,1.3) {$2$};
\node[font=\normalsize,left] at (0.1,0) {$3$};
\end{tikzpicture}
\par\normalsize Comparator $T^*$
\end{minipage}\hfill
\begin{minipage}[t]{0.23\textwidth}
\centering
\begin{tikzpicture}[x=0.64cm,y=0.64cm]
\path[use as bounding box] (-0.5,-0.4) rectangle (3.5,4.3);
\node[keynode] (a4) at (1.5,3.9) {$4$};
\node[keynode] (a1) at (0,1.9) {$1$};
\node[keynode] (a2) at (1.5,1.9) {$2$};
\node[keynode] (a3) at (3,1.9) {$3$};
\draw[ed] (a4)--(a1) (a4)--(a2) (a4)--(a3);
\node[font=\normalsize] at (0,1.25) {$1$};
\node[font=\normalsize] at (1.5,1.25) {$3$};
\node[font=\normalsize] at (3,1.25) {$2$};
\node[font=\normalsize] at (0,0.65) {T};
\node[font=\normalsize] at (1.5,0.65) {T};
\node[font=\normalsize] at (3,0.65) {Q};
\end{tikzpicture}
\par\normalsize Old auxiliary heap
\end{minipage}\hfill
\begin{minipage}[t]{0.23\textwidth}
\centering
\begin{tikzpicture}[x=0.64cm,y=0.64cm]
\path[use as bounding box] (-0.5,-0.4) rectangle (3.5,4.3);
\node[keynode] (b4) at (1.5,3.9) {$4$};
\node[keynode] (b1) at (0.2,2.2) {$1$};
\node[keynode] (b3) at (2.7,2.2) {$3$};
\node[keynode] (b2) at (1.5,0.4) {$2$};
\draw[ed] (b4)--(b1) (b4)--(b3);
\draw[shorted] (b3)--(b2);
\end{tikzpicture}
\par\normalsize After \emph{short links}
\end{minipage}\hfill
\begin{minipage}[t]{0.23\textwidth}
\centering
\begin{tikzpicture}[x=0.64cm,y=0.64cm]
\path[use as bounding box] (-0.5,-0.4) rectangle (3.5,4.3);
\node[keynode] (c4) at (2,3.9) {$4$};
\node[keynode] (c1) at (0.5,2.6) {$1$};
\node[keynode] (c3) at (1.7,1.3) {$3$};
\node[keynode] (c2) at (0.5,0) {$2$};
\draw[ed] (c4)--(c1);
\draw[longed] (c1)--(c3);
\draw[shorted] (c3)--(c2);
\end{tikzpicture}
\par\normalsize After \emph{long links}
\end{minipage}
\caption{The comparator, the old auxiliary heap, and the two steps of
geometric assembly for the four-key access. Labels outside the vertex
circles give comparator ranks. The old children, in key order, are two touched singletons
(T-atoms) and one untouched gap representative (a Q-atom). The dashed
edge is the \emph{short link} $3\to2$; the thick edge is the \emph{long link} $1\to3$.
Thus key $3$ wins a child and then loses to $1$. The comparator happens
to coincide with the final heap in this example; they have different
definitions and need not coincide in general.}
\label{fig:assembly}
\end{figure}

The middle assembly step illustrates the geometry; it need not be the
canonical heap of either endpoint height profile. It introduces no
accounting update: \cref{sec:barriers} specifies the single frozen
snapshot used for the entire access. A key can also win and lose
within the same long-link forest.

\subsubsection{Interval cuts and signed crossings}\label{sec:workedcut}
A cut is a vertical dividing line between consecutive keys. The next lemma explains why many links can cross the same cut while having a small \emph{signed} total contribution. We will apply it to the long-link Cartesian trees; its accounting applications begin in \cref{sec:variation}.

\begin{lemma}[Ordered-tree cut lemma]\label{lem:cutpath}
Let $T$ be a BST on an ordered representative list, let $c$ be a cut between keys, and let $Q$ be a consecutive sublist. Orient edges from parent to child: the endpoints of an edge $v\to w$ are its parent $v$ and child $w$.
\begin{enumerate}[label=(\roman*)]
\item Edges crossing $c$ form a subsequence of the search path for $c$; their crossing directions alternate. They need not be the entire path.
\item If $c$ separates two members of $Q$, crossing edges whose parent and child both lie in $Q$ form a suffix of the crossing sequence. In particular,
\begin{equation}\label{eq:signedcut}
 \left|\sum_{v\to w\text{ crossing }c,\ v,w\in Q}\sgn(w-v)\right|\le1.
\end{equation}
\item If $c$ is immediately outside one end of $Q$, crossings with an endpoint in $Q$ form a suffix. When ranks strictly increase down $T$, their inside-endpoint ranks are nondecreasing, allowing repeated endpoints.
\item For any selected edges on a search path with strictly increasing ranks, the parent-to-child rank intervals are increasingly ordered and have disjoint interiors.
\end{enumerate}
\end{lemma}
\begin{proof}
A subtree from which the search has diverged is entirely on one side of $c$, so it contains no crossing edge. Each crossing changes sides; hence the directions alternate. After a crossing whose two endpoints belong to $Q$, these endpoints give lower and upper search bounds in $Q$. All subsequent path keys lie between them, proving (ii). At the lower boundary of $Q$, the first inside endpoint gives an upper bound in $Q$; all later endpoints above $c$ stay in $Q$. The upper-boundary case is symmetric, proving (iii). Successive occurrences of inside endpoints follow the path order, with a possible repetition at two incident edges. Finally, strict rank increase along the path proves (iv).
\end{proof}
The consecutive-block restrictions in the lemma are essential: arbitrary subsets of crossing edges need not have the suffix property.

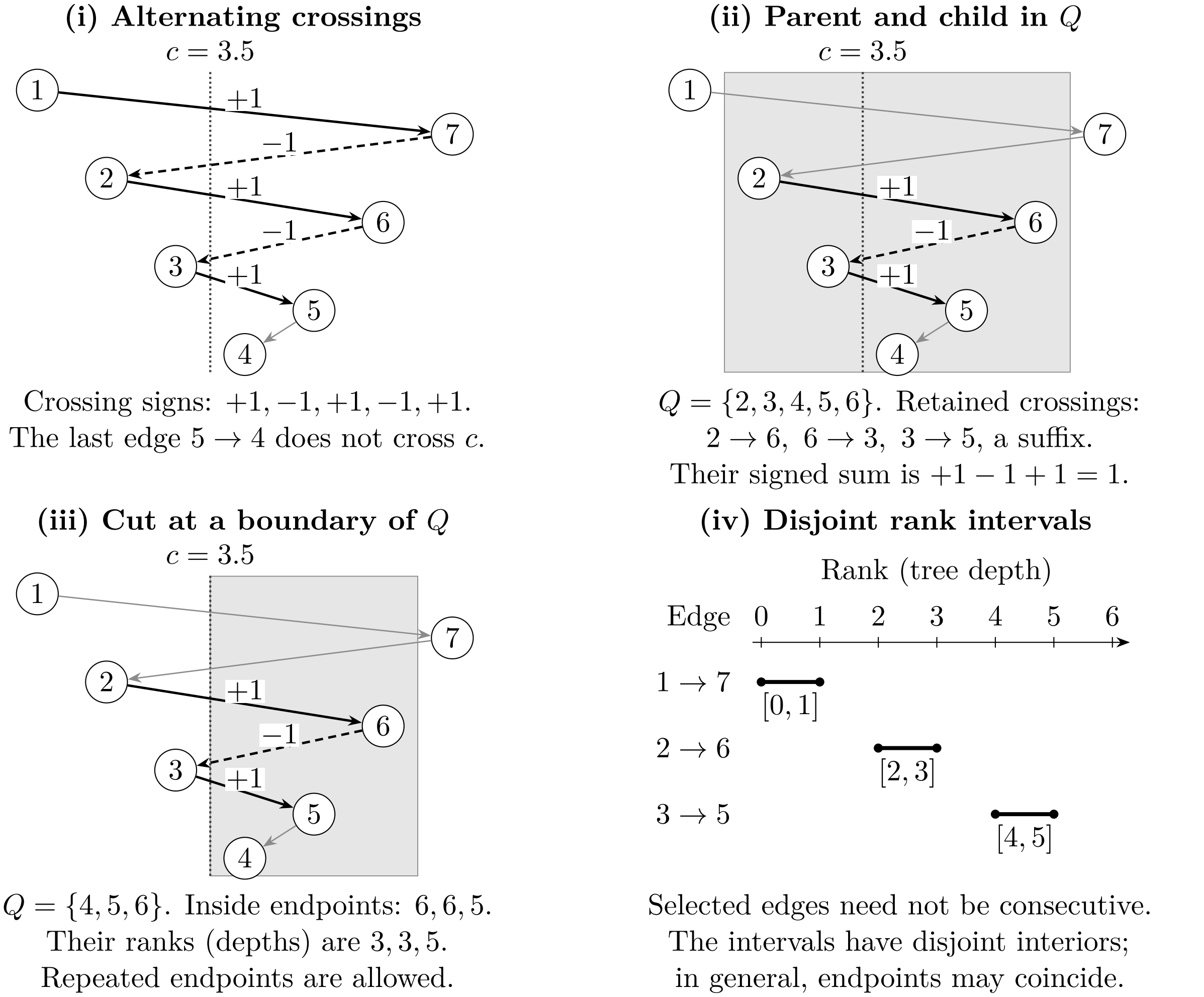
\begin{figure}[!htbp]
\centering
\begingroup
\tikzset{cutnode/.style={circle,draw,fill=white,minimum size=5.5mm,
 inner sep=0pt,font=\normalsize},
 cutedge/.style={-{Stealth[length=1.8mm]},line width=0.9pt},
 cutright/.style={draw=black,solid},
 cutleft/.style={draw=black,dash pattern=on 3pt off 2pt},
 cutexcluded/.style={draw=black!45,solid,line width=0.5pt},
 cutline/.style={draw=black!75,densely dotted,line width=0.8pt},
 cutregion/.style={fill=black!10,draw=black!45,line width=0.35pt}}
\newcommand{\cutbase}{%
 \path[use as bounding box] (0.6,-0.45) rectangle (7.4,7.05);
 \draw[cutline] (3.5,-0.4)--(3.5,6.45);
 \node[anchor=south] at (3.5,6.45) {$c=3.5$};
 \foreach \v/\y in {1/6,7/5,2/4,6/3,3/2,5/1,4/0}
   {\node[cutnode] (v\v) at (\v,\y) {$\v$};}}
\newcommand{\cutlink}[4]{%
 \draw[cutedge,#3] (v#1)--node[midway,above,fill=white,
 inner sep=0.4pt,text=black] {$#4$} (v#2);}
\newcommand{\cutsuffix}{%
 \draw[cutedge,cutexcluded] (v1)--(v7);
 \draw[cutedge,cutexcluded] (v7)--(v2);
 \cutlink{2}{6}{cutright}{+1}
 \cutlink{6}{3}{cutleft}{-1}
 \cutlink{3}{5}{cutright}{+1}
 \draw[cutedge,cutexcluded] (v5)--(v4);}
\begin{minipage}[t]{0.48\textwidth}
\centering
\textbf{(i) Alternating crossings}\par\smallskip
\begin{tikzpicture}[x=0.91cm,y=0.58cm,font=\normalsize]
\cutbase
\cutlink{1}{7}{cutright}{+1}
\cutlink{7}{2}{cutleft}{-1}
\cutlink{2}{6}{cutright}{+1}
\cutlink{6}{3}{cutleft}{-1}
\cutlink{3}{5}{cutright}{+1}
\draw[cutedge,cutexcluded] (v5)--(v4);
\end{tikzpicture}\par
Crossing signs: $+1,-1,+1,-1,+1$.\\
The last edge $5\to4$ does not cross $c$.
\end{minipage}\hfill
\begin{minipage}[t]{0.48\textwidth}
\centering
\textbf{(ii) Parent and child in $Q$}\par\smallskip
\begin{tikzpicture}[x=0.91cm,y=0.58cm,font=\normalsize]
\path[cutregion] (1.5,-0.4) rectangle (6.5,6.4);
\cutbase
\cutsuffix
\end{tikzpicture}\par
$Q=\{2,3,4,5,6\}$. Retained crossings:\\
$2\to6,\ 6\to3,\ 3\to5$, a suffix.\\
Their signed sum is $+1-1+1=1$.
\end{minipage}

\medskip
\begin{minipage}[t]{0.48\textwidth}
\centering
\textbf{(iii) Cut at a boundary of $Q$}\par\smallskip
\begin{tikzpicture}[x=0.91cm,y=0.58cm,font=\normalsize]
\path[cutregion] (3.5,-0.4) rectangle (6.5,6.4);
\cutbase
\cutsuffix
\end{tikzpicture}\par
$Q=\{4,5,6\}$. Inside endpoints: $6,6,5$.\\
Their ranks (depths) are $3,3,5$.\\
Repeated endpoints are allowed.
\end{minipage}\hfill
\begin{minipage}[t]{0.48\textwidth}
\centering
\textbf{(iv) Disjoint rank intervals}\par\smallskip
\begin{tikzpicture}[x=0.77cm,y=0.58cm,font=\normalsize]
\path[use as bounding box] (-1.7,-0.45) rectangle (6.35,7.05);
\node at (3,6.5) {Rank (tree depth)};
\draw[-{Stealth[length=1.8mm]}] (-0.15,4.9)--(6.3,4.9);
\foreach \r in {0,...,6}{
 \draw (\r,4.78)--(\r,5.02);
 \node[above] at (\r,5.05) {$\r$};}
\node[anchor=east] at (-0.35,5.45) {Edge};
\foreach \a/\b/\y/\v/\w in {0/1/4/1/7,2/3/2.5/2/6,4/5/1/3/5}{
 \node[anchor=east] at (-0.35,\y) {$\v\to\w$};
 \draw[black,line width=1.5pt] (\a,\y)--(\b,\y);
 \fill[black] (\a,\y) circle (1.7pt);
 \fill[black] (\b,\y) circle (1.7pt);
 \node[below] at ({(\a+\b)/2},\y) {$[\a,\b]$};}
\end{tikzpicture}\par
Selected edges need not be consecutive.\\
The intervals have disjoint interiors;\\
in general, endpoints may coincide.
\end{minipage}
\endgroup
\caption{The four parts of \cref{lem:cutpath}. Node labels are keys, and
arrows point from parent to child. Solid black arrows marked $+1$
are counted rightward crossings; dashed black arrows marked $-1$ are
counted leftward crossings. Thin gray arrows are excluded. The dotted
vertical line marks the cut $c$, and the outlined light-gray region
marks $Q$. In panels~(iii)--(iv), tree depth is used as an example of
ranks increasing down the tree.}
\label{fig:orderedcut}
\end{figure}

\Cref{fig:orderedcut} illustrates all four parts of the lemma on one
tree. The edge $7\to2$ crosses the cut but is excluded in panel~(ii)
because $7\notin Q$; the edge $5\to4$ has both endpoints in $Q$ but
does not cross the cut. The eight-key access below realizes the displayed
tree as its long-link tree.

\paragraph{Eight-key access example.}
Four keys suffice to show absorption, but not several alternating
crossings of the same cut. For that purpose, take eight keys with
initial BST the right spine $1,2,\ldots,8$, so $h(v)=1-v$, and request
$8$. Every key is touched. Choose the comparator with root $8$ and
the single descending path
\[
 8\to1\to7\to2\to6\to3\to5\to4.
\]
This is a BST: successive keys alternate between the two current search
bounds. Its depth vector, in key order, is
\[
 (r(1),\ldots,r(8))=(1,3,5,7,6,4,2,0).
\]
All seven old auxiliary children of $8$ are T-atoms. There are no short
links; the long-link tree is the displayed path with its first edge
$8\to1$ omitted. Thus it is exactly the seven-key tree in
\cref{fig:orderedcut}, with each depicted tree depth increased by one
to obtain its comparator rank. The edge $8\to1$ attaches the surviving
representative to the request root and is retained, not charged as a link.
Key $2$, for instance, both loses to $7$ and wins $6$ in the same long
stage.

The crossing directions give $+1-1+1-1+1=1$, as in
\cref{lem:cutpath}. Counting crossings without their signs would give
five. Restricting to the consecutive representatives $\{2,3,4,5,6\}$
retains the last three crossings, whose signs are again $+1,-1,+1$.
This is the suffix property that will control historical charges.

\paragraph{Structural consequences.}
The access cost is now at most three plus the number of short and long links. Everything inside each old gap stays intact. What is still missing is an amortized bound on those links while the comparator and its ranks evolve.
\subsection{Block histories and normalized gaps}\label{sec:blocks}
Measuring every child by its absolute comparator depth would make a cheap comparator rotation look expensive: a whole subtree can shift. A block groups consecutive children that can absorb such a shift together. The block partition is chosen by the analysis, but its history cannot be discarded and rebuilt for free.

At completed update states, for each auxiliary node and each side,
partition its same-side child list into consecutive \emph{shrinking
blocks} and one outermost \emph{regular growing block}, which may be
empty. A growing block receives new children only at its outer end.
A shrinking block receives no entries and never becomes growing again.
It may lose members, be split at a key cut, or be transferred as a
whole. Blocks are never merged.

The accounting uses three kinds of complete updates, called
\emph{events}: source-block conversion, access assembly, and comparator
rotation with its block maintenance. \Cref{sec:barriers} specifies
their order and when labels and membership records are evaluated.

During a comparator event, additional temporary singleton growing
blocks are allowed. Each is converted to shrinking before the completed
update state, so there each side again has only its regular
growing block together with its shrinking blocks.

At every time, a key belongs to at most one current block. At each completed update state, every nonroot key belongs to exactly one block and the root belongs to none. Each membership has a single contiguous lifetime and one entry record. Removal ends that lifetime before a replacement entry begins. A split or shrinking transfer preserves the membership record.

\subsubsection{Point gaps and membership records}
For a child $v$ of $u$, define its raw gap $d(v)=r(v)-r(u)$. A growing block has baseline zero. A nonempty shrinking block $Q$ has baseline
\[
 \beta_Q=\min_{v\in Q}d(v).
\]
The \emph{point gap} is $p(v)=d(v)-\beta_Q$; set $p=0$ at the root. At a completed update state its primary label and potential are
\begin{equation}\label{eq:potential}
 g(v)=F(p(v)),\qquad
 \Psi=\sum_{v=1}^n(g(v)-1),\qquad
 0\le\Psi\le(n-1)(G-1).
\end{equation}
Every point gap is nonnegative. Empty shrinking blocks may be removed
from the current block partition, retaining their identities and history;
they have no baseline and contribute zero potential.

For orientation, subtracting the minimum sends both $(10,11,14)$ and $(11,12,15)$ to $(0,1,4)$. Removing the first coordinate of $(10,11,14)$ gives normalized values $(0,3)$ instead of the former $(1,4)$: restriction can only decrease the surviving point gaps. The exact identities are as follows.

For a nonempty set of keys $Q\subseteq[n]$, let
$d=(d(v))_{v\in Q}$ be a vector with one real-valued coordinate for each
key in $Q$. In our application, $Q$ is a shrinking block or one of its
fragments, and $d(v)$ is the raw gap of key $v$. Define the normalized
vector coordinatewise by
\[
 \Norm_Q(d)(v)=d(v)-\min_{w\in Q}d(w)\qquad(v\in Q).
\]
For a uniform shift $c\in\mathbb R$, define $d'(v)=d(v)+c$ for every
$v\in Q$. For a nonempty subset $S\subseteq Q$, write
$d|_S=(d(v))_{v\in S}$ for the vector restricted to the keys in $S$.
The normalization identities for uniform shifts and restrictions are
\begin{equation}\label{eq:normalization}
\begin{aligned}
 \Norm_Q(d')(v)&=\Norm_Q(d)(v)\qquad(v\in Q),\\
 \Norm_S(d|_S)(v)&=\Norm_Q(d)(v)-\min_{w\in S}\Norm_Q(d)(w)\\
                &\le\Norm_Q(d)(v)\qquad(v\in S).
\end{aligned}
\end{equation}
Both follow by subtracting the relevant minima; also
$\Norm_{\{v\}}(d)(v)=0$. The shift identity requires the same value of $c$
for every key in the piece under consideration. The rotation geometry
below is what supplies that requirement.

A shrinking block has point-gap vector $\Norm_Q(d)$. Equation~\eqref{eq:normalization} shows that a uniform raw-gap shift leaves this vector unchanged, while splitting or deleting members can only decrease surviving coordinates. Conversion subtracts a nonnegative minimum from a growing block, and appending to a growing block changes no existing coordinate.

The \emph{current owner} of a block is the auxiliary parent of its
members. This differs from the visibility center $\operatorname{vc}(v)$: a key that is
itself a recursive center has $\operatorname{vc}(v)=v$, whereas its block, when it has
one, is owned by its parent. A block's \emph{original creator} is the key that owned it when it was first opened as growing. These can differ after a shrinking transfer. When a key enters a growing block, its membership records the entry time and the creator's then-current primary label. Those stored data are not updated when the key's or creator's label later changes. The formal accounting will use precisely these records.

An original growing block and all its descendants under splitting form a \emph{root-block lineage}. Its original creator is immutable. While growing, a block has its creator as owner and does not transfer. Its growing period is a single interval of time. After conversion, transfers may change the current owner without changing the original creator.

\Needspace{14\baselineskip}
\paragraph{Terminology.}
The following distinctions will be used throughout the accounting.
\begin{center}
\begin{tabularx}{\textwidth}{@{}p{0.22\textwidth}X@{}}
\toprule
Term & Meaning \\
\midrule
Block / fragment & A consecutive group of same-side auxiliary children; a fragment is a surviving portion of a shrinking block after splits or deletions. \\
Block instance & One block identity counted in $L$ (\cref{lem:births}); a split retires its source and creates new instances. \\
Root-block lineage & An original growing block together with all its split descendants, sharing one original creator and entry order. \\
Membership / record & One key's uninterrupted membership, with immutable entry time and stored creator label; splits and shrinking transfers preserve it. \\
Owner / creator & The current auxiliary parent of the members / the key that owned the original growing block. \\
\bottomrule
\end{tabularx}
\end{center}

\paragraph{Entry times and cuts.}
For a key $v$ in a root-block lineage, its \emph{original entry time}
$\tau_v$ (also called its \emph{original join time}) is the time when
it entered the original growing block. Splitting or transferring the
block preserves this recorded time. After conversion to shrinking, a
\emph{fragment} is a surviving piece of that block, possibly the whole
block if it has not been split.

Choose a time $\tau$ distinct from all entry times. An \emph{original
join-time cut} at $\tau$ divides the lineage's keys into those that
entered before $\tau$ and those that entered after $\tau$. A fragment
\emph{straddles} this cut if it still contains at least one key from
each group.

Give each lineage a fixed sign $\eta_Q$, equal to $+1$ if it originally grows on its creator's right and $-1$ otherwise. If $z$ enters before $w$, then
\begin{equation}\label{eq:joinorder}
 \eta_Q(w-z)>0.
\end{equation}
Thus original entry order is increasing key order for a right growing
block and decreasing key order for a left growing block. Once
shrinking, a fragment contains a subsequence of this unchanged entry
order. We store its elements in key order; historical order need not
agree with its current owner's outward order.

For example, suppose keys enter a right growing block in the order
$2,4,7,9$. A time cut between the entries of $4$ and $7$ separates
$\{2,4\}$ from $\{7,9\}$, just as the key cut $c=5\frac12$ does.
The surviving fragment $\{4,7,9\}$ straddles this cut, whereas
$\{7,9\}$ does not. The key cut stays fixed even if some of these keys
later leave the lineage.

\begin{lemma}[Entry order and key cuts]\label{lem:historycuts}
Fix a root-block lineage and a time $\tau$ during its growing period,
distinct from all entry times.
\begin{enumerate}[label=(\roman*)]
\item There is a fixed key cut $c$ separating the keys that entered
before $\tau$ from those that entered after $\tau$. The same cut
separates these two groups within every surviving fragment. Earlier
entrants lie to the left of $c$ for an originally right growing block,
and to the right of $c$ for an originally left growing block.
\item At any fixed time, at most one current block or fragment of the
lineage contains keys from both groups, and hence straddles the cut.
\end{enumerate}
Both statements remain valid when a fragment changes its owner or
its side relative to that owner.
\end{lemma}
\begin{proof}
For (i), \eqref{eq:joinorder} makes original entry order strictly
monotone in key order. Choose $c$ between the two groups, or outside
the entire key range if one group is empty. Splits and removals only
restrict the set of keys in each fragment, so the same $c$ continues
to separate its earlier and later entrants. Transfers preserve the
keys, original entry times, and original orientation.

For (ii), while the block grows there is only one current block in
the lineage. After conversion, consider each fragment's span from
its earliest to its latest position in the original entry order.
A split at a key cut separates this span into disjoint spans, since
key order agrees with entry order up to reversal. Removals can only
shrink the spans, and transfers leave them unchanged. Shrinking
fragments receive no entries and are never merged. Thus the spans
of current fragments remain disjoint, so at most one can contain
positions on both sides of the chosen time cut.
\end{proof}
The second claim concerns one moment in the execution. A split can
replace a straddling fragment by a smaller straddling descendant, so
successive fragments may straddle the same cut at different times.
The later accounting combines this lemma with complete flushing
(\cref{lem:flush}), which leaves at most one member in each old source
block after an access, to bound contributions over the entire history.

\subsubsection{Update states and frozen labels}\label{sec:barriers}
We evaluate primary labels after complete updates: a source-block
conversion, a full access assembly, or a comparator rotation with its
block maintenance. Immediately before and after every request or
comparator rotation, the auxiliary tree is the canonical heap
$\HH(h,r)$. Conversion changes only blocks and labels, not the heap.

An access uses two accounting updates. First convert the requested
root's growing blocks to shrinking, open their empty replacements,
and calculate the resulting labels. Then freeze all labels, source
memberships, and creator records for the \emph{complete} access forest.
Both short and long links read this same snapshot.

Determine the forest by \cref{lem:assembly}. Remove each loser's old
membership before creating its new one. Enter the loser into its
winner's growing block, recording that winner's frozen label. Order
the entries outward on each side, with the short entry before the
long entry when both occur. Only after all entries and parent changes
are complete do we calculate the new primary labels. The diagram
after short links in \cref{fig:assembly} explains the geometric
construction; it is not a separate accounting state and has no new
label snapshot. No comparator operation is interleaved with this access.

A key may acquire children and change its own parent in the same
forest. Its children still record its single frozen label. Changing
its own membership does not change its identity as the creator of its
children's growing blocks. Each loser appears once in the whole
forest: short losers have been removed from the residual list on
which long links are built.

For precise historical accounting, give entries distinct micro-times
in the prescribed order, with no label change during the entries.
Removal precedes replacement entry, and all label changes follow
completion of the memberships. Thus each label change belongs to at
most one current membership of that key. Comparator events use the
same frozen-record convention, with the maintenance specified in
\cref{sec:rotation}.

\subsubsection{Block initialization and access updates}
Initially create two regular growing blocks per key and insert each
nonroot key into the appropriate block of its parent, in outward order.
This creates $2n$ blocks and $n-1$ memberships. Record the creator
labels from the completed initial configuration, keeping those labels
fixed during the initialization entries.

The root conversion opens two new empty regular growing blocks, one
on each side, and leaves all frozen source blocks shrinking. The access
forest appends to the destination growing blocks
in the outward order of \cref{lem:assembly}. An access link needs no
destination conversion, split, or new block. In particular, even a
winner that loses its own membership retains its growing child blocks.

\begin{lemma}[Complete flushing]\label{lem:flush}
After a complete access and its membership update, every old source
block has at most one member. A shrinking block can therefore
participate in at most one nonsingleton flush; afterwards it is empty
or a singleton and can suffer at most one further loss. Only source
blocks owned by the accessed root participate in internal links.
\end{lemma}
\begin{proof}
On each side, only the Cartesian root remains a child of the accessed
key. Every other old child changes parent. A source block lies on
one side and receives no entries. The last claim follows because only
old children of the accessed root change parents in \cref{lem:assembly}.
\end{proof}
Each access creates exactly two new empty regular growing blocks at
the requested root, regardless of the number of links.

\subsubsection{Four-key block history}\label{sec:workedblocks}
Return to the four-key access in \cref{sec:workedassembly}, with ranks
$r=(1,3,2,0)$. Initialize its auxiliary heap as prescribed above.
The three children of $4$ form one growing left block $Q_0$. They
enter in outward order $3,2,1$, although tables list keys in increasing
order. Their initial creator records all equal $g(4)=1$.

\begin{table}[htbp]
\centering\normalsize
\begin{tabular}{@{}llccc@{}}
\toprule
State & Members in $Q_0$ & $\beta_{Q_0}$ &
$(p(1),p(2),p(3))$ & $(g(1),g(2),g(3))$ \\
\midrule
Initial, growing & $\{1,2,3\}$ & $0$ & $(1,3,2)$ & $(2,3,3)$ \\
After conversion & $\{1,2,3\}$ & $1$ & $(0,2,1)$ & $(1,3,2)$ \\
After full access & $\{1\}$ & $1$ & $(0,1,1)$ & $(1,2,2)$ \\
\bottomrule
\end{tabular}
\caption{The accounting states of the four-key access. Both short
and long links use the middle row's frozen labels. The final row
uses each key's new parent and block. The root has $p(4)=0$, $g(4)=1$.
There is no label update between the two geometric link constructions.}
\label{tab:fourlabels}
\end{table}

Conversion subtracts the minimum raw gap, $1$, without changing any
parent. Freeze the resulting labels $(1,3,2)$. The short link
$3\to2$ enters $2$ into the growing left block of $3$, recording
the frozen value $g(3)=2$. The long link $1\to3$ enters $3$ into
the growing right block of $1$, recording the frozen value $g(1)=1$.
Only after both entries are complete do we recompute gaps and labels.
Both losers have final gap $1$, so the labels become $(1,2,2)$.
The block containing $2$ remains attached to $3$ when $3$ changes
parent; neither its membership nor its stored value is reset.

Both links are internal with respect to the same frozen source block.
Key $2$ decreases from label $3$ to $2$, but key $3$ remains at label
$2$. Thus even this small example requires more than immediate
potential drops to pay for every link. The old block $Q_0$ retains
only key $1$ and cannot receive further members. Its creator remains
$4$; the two new memberships have creators $3$ and $1$.

\paragraph{Structural consequences.}
An access creates two new growing blocks at the requested root and
flushes each old source block to at most one member. We next show
that comparator rotations preserve the same block invariants at
controlled cost.

\subsection{Comparator rotations and block maintenance}\label{sec:rotation}
A comparator rotation changes comparator depths and may change which keys represent auxiliary intervals. We need both numerical locality (uniform shifts on each piece of a block) and geometric locality (each transferred piece stays consecutive under one parent). Neither property alone suffices. The fixed-cut schedule below proves them for every valid incoming block partition.

\subsubsection{Interval minima under rotations}
Consider a right rotation in the comparator $T^*$, promoting $a$ above
its parent $b$. Here $a$ is the left child of $b$, so $a<b$. We call $a$ and $b$ the
two \emph{pivot keys} of the rotation. After the rotation, $b$ is the
right child of $a$. Let $d$ be the depth of $a$ in the comparator
immediately before the rotation. Left rotations are symmetric.

\begin{lemma}[Changing minimum pair]\label{lem:rotationmin}
On every fixed key interval, the minimum-depth comparator key stays unchanged or changes from $b$ to $a$.
\end{lemma}
\begin{proof}
Let $I$ be a fixed nonempty key interval, and let $S_b$ be the set of
keys in the old comparator subtree rooted at $b$. Only keys in $S_b$
can change depth during the rotation.

Suppose $I$ contains a strict ancestor $u$ of $b$. The depth of $u$
stays unchanged, and every key in $S_b$ remains strictly below $u$
after the rotation. Since $u\in I$, no key in $S_b$ can be the
minimum-depth key of $I$, either before or after the rotation.
The minimum therefore comes from $I\setminus S_b$, where all depths
are unchanged, so the minimum key stays the same.

An interval intersecting $S_b$, but not contained in $S_b$, contains
such a strict ancestor of $b$. To see this, choose
$x\in I\cap S_b$ and $y\in I\setminus S_b$. Their lowest common
ancestor in the old comparator is a strict ancestor of $b$. Its key
lies between $x$ and $y$, so it belongs to the key interval $I$.
The preceding argument therefore applies. If $I$ is disjoint from
$S_b$, all its keys keep their depths, and its minimum is also unchanged.

It remains to consider $I\subseteq S_b$. If $I$ contains both pivot
keys, its minimum changes from $b$ to $a$. If it contains exactly one
pivot, that pivot remains its minimum. If it contains neither, the
interval lies entirely in one of the three old comparator subtrees:
the left subtree of $a$, the right subtree of $a$, or the right
subtree of $b$. Depths in these subtrees respectively decrease by one,
stay unchanged, or increase by one. Each shift is uniform, so the
minimum key within $I$ stays unchanged.
\end{proof}

\paragraph{Effect on the auxiliary heap.}
The Greedy heights remain fixed during this comparator rotation and
the corresponding auxiliary-heap update. By
\cref{lem:rotationmin}, a recursive interval can change its center only
from $b$ to $a$. If no such interval exists, all recursive choices and
visible sets remain the same, so the auxiliary topology is unchanged.
The following lemma describes the only possible topological change.

\Needspace{26\baselineskip}
\begin{lemma}[Locality of a comparator rotation]\label{lem:rotationlocality}
Suppose a recursive interval $J$ changes center from $b$ to $a$.
Let $J_a$ be the old auxiliary subtree interval rooted at $a$, and
let $J'_b$ be the new auxiliary subtree interval rooted at $b$.
Call a recursive gap heap \emph{exposed at the two pivots} if its root
is a direct auxiliary child of $a$ or $b$ and it contains neither pivot.
Then the following properties hold.
\begin{enumerate}[label=\textup{(\roman*)}]
\item Both recursive constructions reach the same interval $J$.
The key $a$ is a direct old auxiliary child of $b$, and $b$ is a
direct new auxiliary child of $a$.
\item The set
\begin{equation}\label{eq:starunion}
 S=\{v\in J:\operatorname{vc}_{\mathrm{old}}(v)\in\{a,b\}\}
  =\{v\in J:\operatorname{vc}_{\mathrm{new}}(v)\in\{a,b\}\}
\end{equation}
is invariant. The maximal intervals of $J\setminus S$ are the exposed
gap heaps, which retain all their internal edges.
\item Treat each exposed gap heap as one atomic interval and each key
of $S\setminus\{a,b\}$ as a singleton atom. A nonpivot atom has old
parent $a$ exactly when its interval is contained in $J_a$, and new
parent $b$ exactly when its interval is contained in $J'_b$; otherwise
its old parent is $b$ and its new parent is $a$, respectively.
\end{enumerate}
Only the pivots' parent edges and the attachments of these nonpivot
atoms can change.
\end{lemma}
\begin{proof}
All recursive centers above $J$ remain unchanged by
\cref{lem:rotationmin}. Since the heights are fixed, their visible
sets and gap intervals also remain unchanged. Thus both recursive
constructions reach the same interval $J$. By \cref{lem:min}, $J$
lies inside the old comparator subtree of $b$.

The key $a$ is a direct old auxiliary child of $b$. In fact, if it is
visible from $b$, this follows from the heap definition. Otherwise it
lies in a recursive gap of $b$. That gap avoids $b$, lies in its left
comparator subtree, and contains $a$, the root of that subtree; hence
$a$ is its unique minimum-rank key. After the rotation, the symmetric
argument makes $b$ a direct auxiliary child of $a$.

If $a$ is a visible singleton in the old auxiliary heap, then
$J_a=\{a\}$. Similarly,
$J'_b=\{b\}$ if $b$ is a visible singleton in the updated auxiliary heap.

The visibility neighborhoods $N_h(v)$ are fixed. Applying
\cref{lem:rotationmin} to each neighborhood shows that its visibility
center $\operatorname{vc}(v)$ can stay unchanged or move from $b$ to $a$. This statement
concerns visibility centers, not all auxiliary parents. In particular,
the set $S$ in \eqref{eq:starunion} is invariant. It includes both
pivots and the keys assigned to them by
visibility.

The maximal intervals of $J\setminus S$ are exactly the recursive gaps
exposed below the two pivots. If $a$ is a visible singleton immediately
before the comparator rotation, then $S=K(J,b)$ and these are the gaps
of $b$. Otherwise $a$ is the
center of the gap $J_a$ and
$S=K(J,b)\cup K(J_a,a)$: the exposed gaps are the other gaps of $b$
together with those inside $J_a$. After the rotation, the same
description applies with $a$ and $b$ interchanged. Since both $J$ and
$S$ are invariant, the exposed gaps coincide. Each avoids both pivots,
so it lies in one attached comparator subtree. Its heights are fixed
and its ranks shift uniformly; consequently its entire auxiliary heap
is unchanged. Its root may attach to a different pivot afterwards.

Every nonpivot atom attaches directly to $a$ or $b$ in both heaps.
The atoms in the old subtree $J_a$ are precisely those attached to
$a$; the atoms in the new subtree $J'_b$ are precisely those attached
to $b$. This proves the two parent tests. Outside $J$, the recursive
construction is unchanged except that the representative of $J$
changes from $b$ to $a$, completing the locality claim.
\end{proof}

\Needspace{9\baselineskip}
Here and below, a \emph{cut} is a position in key order used to split
blocks. A cut at a half-integer $c$ splits any block with members on
both sides into the members smaller than $c$ and those larger than $c$.
We call each resulting group a \emph{piece of a block}; it is \emph{nonpivot} if
it contains neither $a$ nor $b$. Such a cut splits a block's ordered
list of members, rather than removing a heap edge. Although $J_a$
describes the whole old auxiliary subtree rooted at $a$, its boundary
cuts organize the blocks for the redistribution of children between
the pivots.

\Needspace{10\baselineskip}
\begin{lemma}[Two-interval refinement]\label{lem:cells}
Start with any valid block partition of the auxiliary heap before the
comparator rotation. Split each affected block at the boundaries of
$J_a$ and $J'_b$, and immediately before and after each pivot key
$a,b$. After the rotation and the corresponding auxiliary-heap
update, every resulting piece containing neither pivot consists of
consecutive children of a single parent, all on the same side of that
parent in key order. Adding cuts at rank-shift boundaries makes every
such piece undergo a uniform raw-gap shift.
\end{lemma}
\begin{proof}
First consider a piece produced by these cuts from an old block owned
by $a$ or $b$, containing neither pivot. All members of the piece have
the same parent in the updated auxiliary heap by the parent tests in
\cref{lem:rotationlocality}, because the piece crosses no boundary of
$J'_b$. This common parent may differ from their
old parent. All members also lie on the same side of their new parent
in key order, because the cuts immediately before and after each pivot
prevent a piece from spanning that parent's key.

We next show that the members of the piece are consecutive in their
new parent's child list: no child outside the piece appears between
two of its members. Take two members $v<w$ and a child $z$ of their
new parent with $v<z<w$. Neither pivot can be $z$, because cuts
immediately before and after each pivot isolate that key. If $z$
comes from the same old child list, it belongs to the same old block
because that block consists of consecutive children in that list.
The absence of a cut between $v,w$ then places $z$ in this piece.
If $z$ comes from the other pivot's
old list, its atom has the opposite membership in $J_a$. If
$v,w\in J_a$, the interval property excludes $z\notin J_a$ between
them. If $v,w\notin J_a$ but $z\in J_a$, a boundary cut of $J_a$
separates $z$ from $v$ or $w$, contradicting the absence of a cut in
the piece. Thus every intervening child belongs to the piece.

At the external auxiliary parent, only the representative of the whole
interval $J$ changes, from $b$ to $a$, and cuts immediately before and
after the pivot keys isolate this change. All other auxiliary child
lists are unchanged. If no recursive minimum changes, no topological
transfer is needed.

Finally, in a piece contained in one rank-shift region, all child ranks
change by the same amount. Its members share one old parent and one
new parent, so their raw gaps also change by the same amount. The
reasoning applies to every old block, not only to a maximal same-side
list: a smaller block is a further consecutive subinterval and
inherits the same properties.
\end{proof}

\subsubsection{Fixed-cut maintenance}
The number of moved keys can be large. We maintain the blocks by
cutting them into pieces satisfying three properties: each piece
(a) has all its members attached to the same destination parent and on
the same side of that parent in key order;
(b) remains consecutive in that parent's child list; and
(c) undergoes a uniform raw-gap shift. Properties (a) and (b) preserve
the block partition; property (c) allows minimum subtraction to absorb
the shift. We first specify cuts ensuring these
properties, then give the maintenance procedure and count the new block
instances.

\label{sec:rotationschedule}
If an unchanged auxiliary edge $u\to v$ changes its raw gap, then $u$ is an old comparator ancestor of $a$. Otherwise its whole comparator subtree is disjoint from the rotation or contained in one uniformly shifted attached subtree, so $u,v$ shift equally. The topologically affected owners are also old ancestors of $a$: the pivots and, where present, the auxiliary parent of $b$, which is a strict comparator ancestor of $b$ and hence of $a$ by \cref{lem:ownership}. The pre- and post-rotation auxiliary parents of either pivot also lie on this old ancestor path, by the same ancestor property and the change in comparator ancestor paths under the rotation. Thus all owners involved in resetting pivot memberships are covered as well. There are $d+1$ such ancestors, including $a$.

The standard rotation decomposition has at most five key intervals: the outside-left region, $a$ and its unchanged-side subtree, the middle subtree, $b$ and its unchanged-side subtree, and the outside-right region. Their depth shifts are respectively $0,-1,0,+1,0$, giving at most four boundaries; see also \cite{CHH26}. Relative depth order can change between nonpivot keys in different regions. The argument uses the interval-minimum lemma, not preservation of all pairwise priority comparisons.

\begin{figure}[htbp]
\centering
\begin{tikzpicture}[x=0.85cm,y=0.75cm,
 subtree/.style={draw,isosceles triangle,shape border rotate=90,
 minimum width=9mm,minimum height=9mm,inner sep=1pt,font=\normalsize}]
\node[keynode] (b) at (2.5,3) {$b$};
\node[keynode] (a) at (1.2,1.7) {$a$};
\node[subtree] (A) at (0.3,0) {$A$};
\node[subtree] (B) at (2.1,0) {$B$};
\node[subtree] (C) at (3.8,1) {$C$};
\draw[ed] (b)--(a) (a)--(A) (a)--(B) (b)--(C);
\node[font=\normalsize] at (2,-0.9) {Before};
\draw[-{Stealth},ed] (4.9,1.7)--(6.2,1.7);
\node[keynode] (aa) at (8.2,3) {$a$};
\node[keynode] (bb) at (9.5,1.7) {$b$};
\node[subtree] (AA) at (6.9,1) {$A$};
\node[subtree] (BB) at (8.6,0) {$B$};
\node[subtree] (CC) at (10.4,0) {$C$};
\draw[ed] (aa)--(bb) (aa)--(AA) (bb)--(BB) (bb)--(CC);
\node[font=\normalsize] at (8.7,-0.9) {After};
\node[font=\normalsize,left] at (6.3,1) {$-1$};
\node[font=\normalsize,left] at (8.0,0) {$0$};
\node[font=\normalsize,right] at (11,0) {$+1$};
\end{tikzpicture}
\caption{A right comparator rotation. Every rank in $A\cup\{a\}$
decreases by one, ranks in $B$ stay fixed, and every rank in
$\{b\}\cup C$ increases by one. Outside the rotated subtree ranks
stay fixed. A block spanning two such regions must first be split;
within a resulting piece, a common raw-gap shift cancels when its
minimum is subtracted.}
\label{fig:rotationregions}
\end{figure}
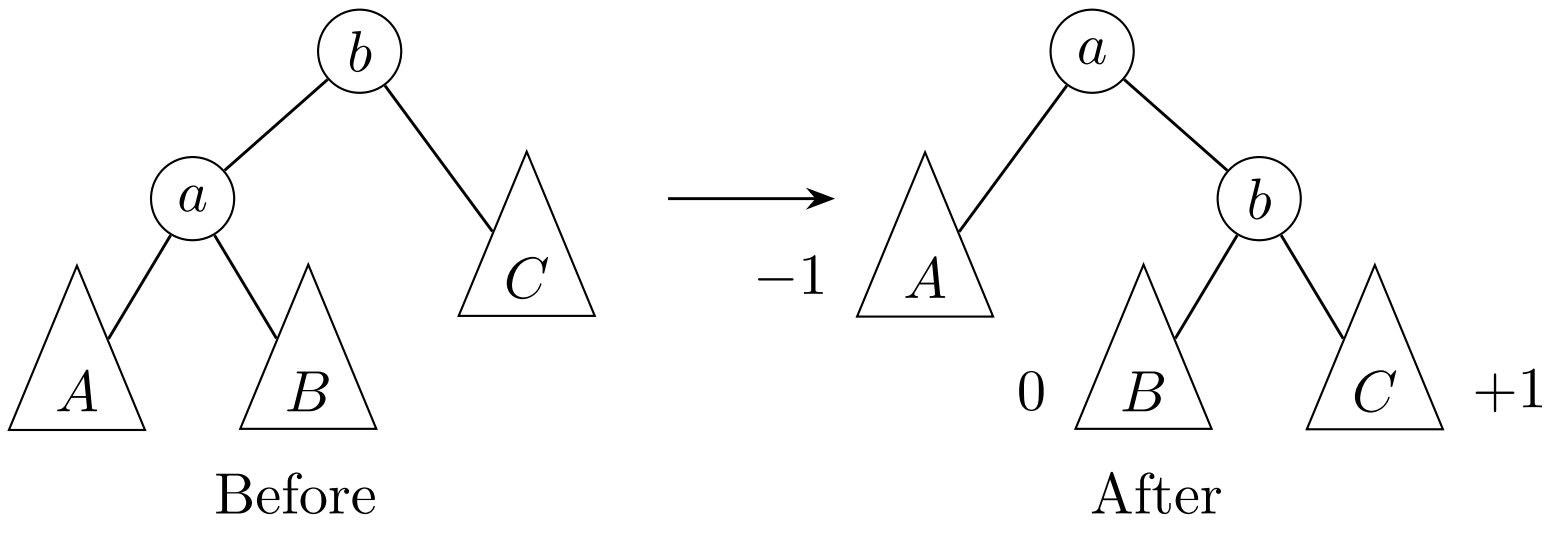

Select the two boundary positions of each of $J_a$ and $J'_b$,
giving four cuts in key order. Select the positions immediately before
and after each of the two pivot keys, giving another four cuts.
Finally, select the boundaries between the at most five depth-shift
regions, giving at most four more cuts. Thus there are at
most $4+4+4=12$ cut positions. Some positions may coincide, and empty
regions may reduce the count; twelve is an upper bound. An interval
$[\ell_J,u_J]$ contributes the half-integer cuts
$\ell_J-\tfrac12$ and $u_J+\tfrac12$, separating keys below, inside,
and above that interval. When no recursive minimum changes, the
boundary cuts of $J_a$ and $J'_b$ can be omitted.

The interval-boundary and pivot cuts ensure properties (a) and (b)
of the pieces---a common parent and side, and consecutive positions
in that parent's child list---by \cref{lem:cells}. The rank-shift
cuts ensure property (c), a uniform raw-gap shift. Apply the following
procedure to the $d+1$ relevant owners.
\begin{enumerate}[label=(\roman*)]
\item Convert each owner's two regular growing blocks to shrinking and
create two new empty regular growing blocks, one on each side of that
owner. Conversion introduces minimum
subtraction, which absorbs a uniform raw-gap shift by
\eqref{eq:normalization}; a growing block's zero baseline would not
absorb it. Conversion also allows subsequent transfers while retaining
the original creator and membership records.
\Needspace{4\baselineskip}
\item Split shrinking blocks at every selected cut they straddle. One
key cut splits at most one block per owner. The pivot cuts isolate the
old pivot memberships; remove each pivot's old membership, if it has one.
\item Transfer every nonpivot piece to its new parent and side,
preserving its records. The three properties above ensure that these
are valid consecutive blocks and that normalization cancels each
piece's uniform raw-gap shift.
\item For each pivot that is nonroot in the updated auxiliary heap,
create a temporary singleton growing block at its position in its
post-rotation parent's child list, enter the pivot, and immediately
convert that block to shrinking without opening a new growing block
in its place. This creates a new membership even if the pivot's
auxiliary parent is unchanged. A pivot at the root after the update
receives no membership. The pivot cuts ensure that no new
singleton is inserted between members of a surviving block. Keep the
new empty regular growing blocks outermost and process temporary blocks one
at a time, so at most three growing blocks with one creator coexist.
\end{enumerate}
Pivot entries use the frozen pre-event labels. Update labels only after
all replacement memberships have been created, so any decrease of an
entering pivot's label to one belongs to its new membership.

A binary split creates two descendants and retires its source. Counting new regular growing blocks, split descendants, and temporary singleton blocks, we obtain the convenient loose bound
\begin{equation}\label{eq:rotationbirths}
 2(d+1)+2\cdot12(d+1)+2\le32(d+1).
\end{equation}
At most two new pivot memberships occur, and all splits use the prescribed cuts.

\begin{lemma}[Comparator monotonicity and block births]\label{lem:births}
At a completed comparator event, no key's primary label
$g(v)=F(p(v))$ increases. If $L$ is the total number of block instances
ever created, including split descendants, then
\begin{equation}\label{eq:L}
 L\le2n+2m+32\EE.
\end{equation}
Comparator events create at most $2R$ new memberships in total.
\end{lemma}
\begin{proof}
Each pivot has point gap zero after the event, either at the root or by singleton normalization. Conversion and refinement restrict every nonpivot to a subset of its old block; they cannot increase its normalized gap. The prescribed cuts ensure a uniform raw-gap shift on each transferred piece, which cancels by \eqref{eq:normalization}. Hence no primary label increases at the completed update state; the comparison uses the completed comparator states.

Initialization creates $2n$ blocks, an access creates two, and \eqref{eq:rotationbirths} pays for each comparator event. Only the two pivots can require replacement membership records. Summation proves the claims.
\end{proof}

\subsubsection{Four-key rotation example}\label{sec:workedrotation}
After the four-key access, hold $h=(1,1,-3,1)$ fixed and rotate the
comparator right at $4$, promoting $1$. This is one comparator step
between accesses; no new Greedy row is added. The comparator changes
from the path $4\to1\to3\to2$ to $1\to4\to3\to2$, with the
appropriate BST left and right edges. Here $A=C=\varnothing$ and $B$
is the subtree on $\{2,3\}$ in \cref{fig:rotationregions}. Consequently
only the pivot ranks change, from $r=(1,3,2,0)$ to $(0,3,2,1)$.

The auxiliary heaps happen to be those same paths. The growing block
$\{3\}$ under $1$ is converted and transferred to $4$ as a shrinking
singleton. Its old point gap is $1$; after normalization its point gap
is $0$. Its original creator is still $1$, and its stored creator
value is still $1$, even though its current owner is now $4$.
Key $2$ stays under $3$ with point gap $1$ and stored value $2$.

\begin{center}\normalsize
\begin{tabular}{@{}ccccp{0.34\textwidth}@{}}
\toprule
Key & Rank, old/new & Parent, old/new & Label, old/new & Record action \\
\midrule
$1$ & $1/0$ & $4/\text{root}$ & $1/1$ & Old pivot membership ends. \\
$2$ & $3/3$ & $3/3$ & $2/2$ & Keep creator $3$, stored value $2$. \\
$3$ & $2/2$ & $1/4$ & $2/1$ & Keep creator $1$, stored value $1$. \\
$4$ & $0/1$ & $\text{root}/1$ & $1/1$ & New singleton membership, stored value $1$. \\
\bottomrule
\end{tabular}
\end{center}

This separates a history-preserving transfer from a new entry. Moving
the nonpivot block $\{3\}$ changes its owner but does not create a
record; the pivots are handled by the explicit reset rule. In larger
blocks the same normalization works for every member of a uniformly
shifted piece. The cuts in \cref{lem:cells} are what make that uniformity
and destination contiguity valid.

\subsubsection{Persistent-history invariant}\label{sec:invariant}
The following proposition connects the structure to the accounting.
It also verifies that combining all access links into one frozen
update preserves the historical properties used in the proof.

\begin{proposition}[Validity of the maintained history]\label{prop:invariant}
For every finite initial height profile and comparator execution,
the preceding maintenance defines one history with these properties:
\begin{enumerate}[label=(\roman*)]
\item Each key belongs to at most one current block; each nonroot key
has exactly one membership at every completed update state.
\item Memberships have contiguous lifetimes and immutable entry records,
preserved by splits and shrinking transfers.
\item Shrinking blocks never receive entries, become growing, or merge.
A growing block keeps its creator and owner throughout its growing period.
\item Original entry order satisfies \eqref{eq:joinorder}, and split
descendants have disjoint spans in that order.
\item At most three growing blocks with the same creator coexist.
\item All short and long links of an access read one frozen snapshot.
Each loser enters one destination once; labels are updated after all
entries, in both access and comparator events.
\end{enumerate}
\end{proposition}
\begin{proof}
Initialize each nonroot key in one parent block, in outward order,
recording labels from the completed configuration. Before an access,
convert the root's growing blocks and freeze the resulting labels
and memberships. Every short loser is absent from the residual
Cartesian trees, and every nonroot of those trees is a single long
loser. Hence each losing key has one destination across the whole
forest, not merely within each geometric part.

Read all old records from the snapshot, remove losing memberships,
and make the new entries in outward order. Sources belong to the
accessed root, whereas destinations belong to its old children, so
they are distinct. By \cref{lem:assembly}, a winner receives at most
one new child of each type per side, with short before long in
outward order. This preserves contiguity, the outermost growing block,
and the entry-order invariant. A winner's simultaneous loss changes
its own membership but not its creator identity or its growing child
blocks. Thus combining the forest changes no structural requirement.

For a comparator event, conversion precedes splitting and transfer.
Fixed-cut refinement preserves the nonpivot records and entry positions;
\cref{lem:cells} ensures destination contiguity. Remove old pivot
memberships before their singleton replacement entries, and convert
the temporary growing blocks one at a time. This preserves uniqueness
and the bound of three growing blocks per creator. No shrinking block
receives entries, regrows, or merges. A growing block loses members
only after conversion, so an ended membership cannot resume. Splits
and deletions preserve disjoint original entry spans.

After access assembly or comparator maintenance the heap is canonical.
All records are created before the final label update. Each label
change therefore belongs to at most one then-current membership; a
key becoming root has no membership and its decrease is unassigned.
Induction over these updates proves all six properties in the same
maintained history.
\end{proof}

\paragraph{Maintenance bounds.}
We have an exact auxiliary representation, complete source-block
flushing, and one valid history with $L\le2n+2m+32\EE$.
Comparator events never increase primary labels. The amortized proof
counts links in precisely this history, using one snapshot per access.

\section{Amortized analysis}\label{sec:amortized}
We analyze the complete access forests using the single snapshot
specified in \cref{sec:barriers}. The proof has three parts:
(a) a shrinking-block population count bounds total block--access incidences;
(b) ordered cuts bound total label variation and weighted boundary
differences; and (c) membership histories and a capped-record potential
bound the number of internal links.
We prove (a) and (b) in \cref{sec:variation} and (c) in
\cref{sec:accounting}. In \cref{sec:closing}, we combine these bounds
to obtain the comparator-relative cost bound.
Throughout, $M=n+m+\EE$ and $G=1+\bitlen(H)$.
\Cref{tab:accounting} collects the accounting quantities; those not
yet defined are introduced below. The $O(MG)$ bounds hold for every
comparator described in \cref{sec:parameters}.

\begin{table}[!t]
\centering
\caption{Accounting quantities and where they are bounded.}
\label{tab:accounting}
\normalsize
\begin{tabularx}{\textwidth}{@{}p{0.15\textwidth}>{\raggedright\arraybackslash}X>{\raggedright\arraybackslash}p{0.34\textwidth}@{}}
\toprule
Symbols & Meaning & Bound and reference \\
\midrule
$R,\EE,M$ & Rotation count; weighted rotation cost; total structural budget
& $R\le\EE\le M$; normalized $M=O(n+\OPT_n(X))$ (\cref{sec:parameters}) \\[0.4em]
$H,G$ & Maximum comparator height; maximum label $G=1+\bitlen(H)$
& Normalized $G=O(\log\log n)$ (\cref{sec:parameters}) \\[0.4em]
$L,\Act$ & Block instances ever created; active block--access incidences
& $L,\Act=O(M)$ (\cref{lem:births,lem:active}) \\[0.4em]
$I,B,\Neq$ & Internal links; boundary links; equal-label internal links
& Each is $O(MG)$ (\cref{lem:boundarybudgets,lem:internalcount}) \\[0.4em]
$P^+,P^-,V$ & Total label increases, decreases, and variation $V=P^++P^-$
& $V=O(MG)$ (\cref{lem:boundarybudgets}) \\[0.4em]
$W^+,W^-,W$ & Positive and negative parts of boundary loser-minus-winner label differences; $W=W^++W^-$
& $W=O(MG)$ (\cref{lem:boundarybudgets}) \\[0.4em]
$\Psi$ & Primary-label potential
& $0\le\Psi\le(n-1)(G-1)$; \eqref{eq:potential} \\[0.4em]
$\Gamma_t$ & Capped creator-record potential
& $0\le\Gamma_t\le(n-1)G$; \eqref{eq:cappedrecord} \\[0.4em]
$J_0$ & Initial and pivot entry budget $J_0=n-1+2R$
& $J_0=O(M)$; \eqref{eq:Tinternal} \\[0.4em]
$\Tint$ & Signed internal record-change sum
& $\Tint\le GJ_0+W^++P^-$ (\cref{lem:capped}) \\
\bottomrule
\end{tabularx}
\end{table}

\subsection{Block participation and boundary estimates}\label{sec:variation}
Classify every short or long link using the source blocks immediately
after the root conversion, before any link of that access. A link is
\emph{internal} when its winner and loser belong to the same source
block, and a \emph{boundary link} otherwise. Let $I,B$ count these
links over the whole execution, and let $\Neq$ count internal links
with equal frozen endpoint labels. The geometric distinction between
short and long links is independent of this block distinction.

\paragraph{Subtraction and label progress.}
We first record the elementary numerical property used throughout.
\begin{lemma}[Bit-length subtraction]\label{lem:bits}
For integers $0\le a<b\le H$, $F(b-a)\le F(b)$. If $F(a)=F(b)$,
then $F(b-a)\le F(b)-1$. If $F(b-a)=F(b)$, then $b>2a$.
\end{lemma}
\begin{proof}
Monotonicity gives the first claim. Label $j\ge2$ corresponds to the
integers in $[2^{j-2},2^{j-1}-1]$, and the difference of two distinct
integers in that interval is less than $2^{j-2}$. This proves the
second claim. For the third, put $F(b)=j$. Equality implies
$b-a\ge2^{j-2}>b/2$, so $a<b/2$.
\end{proof}
The equal-label estimate is the numerical ingredient of~\cite{CHH26}.
Its application here uses the whole-access subtraction identity below.

\begin{lemma}[Point-gap subtraction]\label{lem:dyadic}
For an internal link $v\to w$, let $\beta$ be its source block's
baseline in the frozen access snapshot. Then
\[
 p(v)=r(v)-\beta,\qquad p(w)=r(w)-\beta,\qquad
 p'(w)=r(w)-r(v)=p(w)-p(v),
\]
where $p'(w)$ is the loser's point gap after the complete access.
The loser's label cannot increase, and equal endpoint labels imply a
decrease of at least one.
\end{lemma}
\begin{proof}
The source owner is the comparator root, of rank zero. The destination
is the winner's growing block, with baseline zero. All ranks remain
fixed during the access, and each loser has only one new parent.
Strict comparator ancestry gives $0\le p(v)<p(w)$. Apply
\cref{lem:bits}. The identity uses the original frozen baseline even
when the source block loses short-link members before the long links
are constructed.
\end{proof}

\paragraph{Block--access incidences.}
For access $t$, let $a_t$ count the distinct frozen source blocks
incident to at least one boundary link of its complete forest. Define
\[
 \Act=\sum_t a_t.
\]
A block is counted once in an access, even if it participates in both
short and long boundary links. Both kinds use the same frozen source
blocks, and the per-block estimates below account for their combined
contributions.
\begin{lemma}[Source-block participation]\label{lem:active}
The total number of active block--access incidences satisfies
\begin{equation}\label{eq:active}
 \Act\le4m+32\EE=O(M).
\end{equation}
\end{lemma}
\begin{proof}
Let $Z$ count all nonempty shrinking blocks. Initially $Z=0$, since
all initial blocks are growing. For access $t$, let $N_t$ count its
nonempty frozen source blocks, and let $\Delta_t Z$ be the change
from before root conversion to after complete assembly. Conversion
adds at most two nonempty shrinking blocks. Every source is then
shrinking; assembly leaves at most two of them nonempty, since only
the at most two roots of the access forest remain children of the
accessed root (\cref{lem:assembly}). All link destinations are growing
blocks, and all other shrinking blocks keep their memberships.
Consequently
\[
 a_t\le N_t,\qquad N_t+\Delta_t Z\le4.
\]

A comparator rotation at depth $d$ increases $Z$ by at most
$32(1+d)$. To see this, charge each regular growing-block conversion
to its new empty replacement block, each increase due to splitting
to a new split descendant, and each temporary pivot singleton to its
creation. Transfers preserve the population and deletions cannot increase it.
These charges are covered by the creation bound
\eqref{eq:rotationbirths}.

Summing over accesses and telescoping $Z$ over all events gives
\[
 \Act\le4m-\sum_{t\text{ an access}}\Delta_t Z
 =4m-Z_f+\sum_{t\text{ a rotation}}\Delta_t Z
 \le4m+32\EE,
\]
where $Z_f\ge0$ is the final population.
\end{proof}

\paragraph{Aggregate quantities.}
Index the completed updates by $s=1,\ldots,T$, including conversions,
complete access assemblies, and comparator events. Let $g_s(v)$ be
key $v$'s primary label after update $s$, with $g_0(v)$ its initial
label. Writing $[x]_+=\max\{x,0\}$, define
\begin{align*}
 P^+&=\sum_{s=1}^{T}\sum_{v=1}^{n}[g_s(v)-g_{s-1}(v)]_+,\\
 P^-&=\sum_{s=1}^{T}\sum_{v=1}^{n}[g_{s-1}(v)-g_s(v)]_+,
 \qquad V=P^++P^-.
\end{align*}
Thus $P^+$ is the total amount of label increase and $P^-$ is the
total amount of label decrease; both are nonnegative. The positive
and negative parts are taken before summing coordinates.

For the potential in \eqref{eq:potential}, write
$\Psi_s=\sum_{v=1}^{n}(g_s(v)-1)$, and let $\Psi_0$ and
$\Psi_f=\Psi_T$ denote its initial and final values. Each update
changes the potential by the sum of its label changes:
\[
 \Psi_s-\Psi_{s-1}=\sum_{v=1}^{n}(g_s(v)-g_{s-1}(v)).
\]
Since $[x]_+-[-x]_+=x$, summing over updates cancels the intermediate
potential values and gives
\begin{equation}\label{eq:jordan}
 \begin{aligned}
 P^+-P^-&=\sum_{s=1}^{T}(\Psi_s-\Psi_{s-1})=\Psi_f-\Psi_0,\\
 V&=P^++P^-=2P^++\Psi_0-\Psi_f.
 \end{aligned}
\end{equation}

Measure weighted boundary differences by
\begin{align*}
 W^+&=\sum_{v\to w\text{ boundary}}[g(w)-g(v)]_+,\\
 W^-&=\sum_{v\to w\text{ boundary}}[g(v)-g(w)]_+,\qquad
 W=W^++W^-.
\end{align*}
Every endpoint label is from its complete access's frozen snapshot.
We bound $W$ directly, since replacing each difference by $G$ would
lose an additional factor of $G$.

\Needspace{13\baselineskip}
\begin{lemma}[Boundary and variation budgets]\label{lem:boundarybudgets}
The maintained history satisfies
\begin{align}
 P^+&\le4(G-1)\Act,\label{eq:Pplus}\\
 P^-&\le(n-1)(G-1)+P^+,\qquad
 V\le(n-1)(G-1)+2P^+,\label{eq:Vbound}\\
 B+\Neq&\le P^-+(2G+2)\Act,\label{eq:NBbound}\\
 W^+&\le P^-+4(G-1)\Act,\qquad
 W^-\le P^-+8G\Act.\label{eq:weightedglobal}
\end{align}
In particular, $V$, $B+\Neq$, $W^+$, and $W^-$ are all $O(MG)$,
independently of $I$; hence so is $W=W^++W^-$.
\end{lemma}
\begin{proof}
Fix one access with $a$ active source blocks. For a boundary link
$v\to w$, the final loser label is
$z(w)=F(r(w)-r(v))$. Put
\[
 D^- =\sum_{v\to w\text{ boundary}}[g(w)-z(w)]_+.
\]
Its losers are distinct across the entire short-and-long-link forest,
so these quantities sum to at most $P^-$ over all accesses.

\emph{Label increases.}
Internal losers do not increase by \cref{lem:dyadic}. A surviving
source member has a nondecreasing block minimum, and old destination
members have unchanged raw gaps and baseline zero. Thus only a
boundary loser can increase. If its old source baseline is $\beta_Q$,
its gap changes from $r(w)-\beta_Q$ to $r(w)-r(v)$; an increase
requires $r(v)<\beta_Q$. Call such a boundary loss dangerous.

At most one short edge crosses each endpoint of a source block.
For long edges, the residual members of that block are consecutive
in the residual representative list. At each endpoint, incident
crossings lie along a search path with strictly increasing ranks
(\cref{lem:cutpath}). There is at most one incoming dangerous edge:
after reaching an inside key, of rank at least $\beta_Q$, every later
parent on the path has rank at least $\beta_Q$. Hence there are at
most two short boundary losses and two dangerous long losses per
active block. Each increases a label by at most $G-1$. Comparator
events and conversions do not increase labels, by
\cref{lem:births} and minimum subtraction. This proves
\eqref{eq:Pplus}; \eqref{eq:jordan} proves \eqref{eq:Vbound}.

\emph{Nondecreasing boundary losses.}
For a nondangerous incoming long edge, shift its endpoint ranks by
the loser's source baseline and write
$a_0=r(v)-\beta_Q\ge0$, $b_0=r(w)-\beta_Q>a_0$.
If its label does not decrease, then $F(b_0-a_0)=F(b_0)$, so
$b_0>2a_0$ by \cref{lem:bits}. Along one boundary search path, the
parent rank of the next selected edge is at least the preceding
selected child rank. The selected positive child gaps therefore
strictly double and are bounded by $H$, allowing at most $G-1$
such edges per endpoint. Including at most two dangerous long edges
and two short boundary edges gives at most $2G+2$ exceptional losses
per active block. Every other boundary loss, and every equal-label
internal loss, consumes at least one primary-label decrease. Their
losers are distinct in the complete access, proving \eqref{eq:NBbound}.

\emph{Weighted differences.}
Write $W_t^+,W_t^-$ for the contributions of this access and set
$S_t=W_t^+-W_t^-$. Group
$S_t=\sum_{\rm boundary}(g(w)-g(v))$ by the blocks containing its
endpoints. At a boundary cut of a block $Q$, the incident long
crossings form an alternating suffix of a search path, and the ranks
of their inside endpoints are nondecreasing (\cref{lem:cutpath}(iii)).
Their labels use the same frozen baseline $\beta_Q$, so they too
are nondecreasing. An alternating sum of nondecreasing values in
$[1,G]$ has absolute value at most $G$: expanding into successive
increments gives coefficients equal to suffix sign sums, each of
magnitude at most one. A short edge adds at most one term per cut.
There are two cuts per block, so
\begin{equation}\label{eq:fluxaccess}
 |S_t|\le4Ga.
\end{equation}
Each boundary edge contributes $-g(v)$ at the facing endpoint of its
winner's block and $+g(w)$ at that of its loser's block. Blocks
strictly between its endpoints contribute nothing. This grouping
therefore counts each summand exactly once.

For an outgoing boundary edge with winner in $Q$, put
$a_0=r(v)-\beta_Q\ge0$ and $d=r(w)-r(v)>0$. Monotonicity gives
\begin{align*}
 [g(w)-g(v)]_+
 &\le[g(w)-F(d)]_+ +[F(d)-F(a_0)]_+\\
 &\le[g(w)-z(w)]_+ +F(a_0+d)-F(a_0).
\end{align*}
At each boundary cut of $Q$, outgoing long-edge rank intervals have
disjoint interiors and occur in increasing order along the search
path (\cref{lem:cutpath}(iv)). Their last differences therefore sum
to at most the range $G-1$ of the same monotone function
$r\mapsto F(r-\beta_Q)$. The at-most-one outgoing short edge at
that cut adds at most $G-1$. Over both cuts of every active block,
\[
 W_t^+\le D^-+4(G-1)a,\qquad
 W_t^-=W_t^+-S_t\le D^-+8Ga.
\]
Summing proves \eqref{eq:weightedglobal}. The long-edge calculations
use the original frozen baseline, not a baseline recomputed after
short losses. Restricting the representative list preserves its
consecutive block intervals and does not affect these calculations.

Finally, \cref{lem:active} gives $\Act=O(M)$, and $n\le M$.
All displayed estimates are independent of the unknown internal-link
count $I$.
\end{proof}

\paragraph{Boundary cancellation in the eight-key example.}\label{sec:workedweighted}
Use the long-link path of \cref{sec:workedcut}, with comparator ranks
in path order $r(1),r(7),r(2),r(6),r(3),r(5),r(4)=1,2,3,4,5,6,7$.
For illustration take the frozen shrinking blocks
$Q_1=\{1,2,3\}$ and $Q_2=\{4,5,6,7\}$. This is a separate
admissible snapshot, not a repartitioning during an access. Their
baselines are $1,2$, and their labels in key order are
\[
 (g(1),\ldots,g(7))=(1,3,4,4,4,3,1).
\]
The five boundary edges have signed endpoint differences $0,2,0,1,0$,
so $W_t^+=3$, $W_t^-=0$. Grouping by blocks gives
$-1+3-3+4-4=-1$ for $Q_1$ and $1-1+3-3+4=4$ for $Q_2$.
These are alternating sums of nondecreasing inside labels. Every
boundary loser finishes with label $F(1)=2$, giving $D^-=6$.
The first edge $1\to7$ is dangerous and increases its loser's label;
the aggregate bounds account for it together with the decreases.

\subsection{Membership histories and capped records}\label{sec:accounting}
The boundary estimates leave the internal links whose endpoint labels
differ. Such a link can preserve its loser's label, as the four-key
example demonstrates. We use parts of the historical accounting
argument of~\cite{CHH26}, adding the machinery needed here for Greedy.

Each membership retains the block creator's label at the time of entry
as a fixed record. For an internal link, we compare the winner's current
label with its entry label, compare the creator's labels at the two
endpoints' entry times, and account for the replacement of the loser's
stored record. Together with a signed entry term, these comparisons
give the exact unit-cost identity \eqref{eq:unit}. We first
bound repeated use of label changes and then control record replacements
with a capped potential. The records belong to the maintained history
in \cref{prop:invariant}.

\paragraph{Bounded reuse.}
\begin{lemma}[Winner reuse]\label{lem:wins}
During one root-block membership, a key wins at most four internal links,
including equal-label links.
\end{lemma}
\begin{proof}
Before the first nonsingleton flush of its current fragment it wins
no internal link there. In that access it receives at most one short
and one long child on each side, by \cref{lem:assembly}. Afterwards
it has left the lineage or remains in a shrinking singleton, which
cannot support another internal link. Splits and transfers add no members.
\end{proof}

For an internal link $e:v\to w$, let $\tau_v,\tau_w$ be its
endpoints' original entry times in their old memberships. At a time
$\tau$ distinct from entries define
\[
 \sigma_e(\tau)=
 \begin{cases}
  +1, & \text{if }\tau_v<\tau<\tau_w,\\
  -1, & \text{if }\tau_w<\tau<\tau_v,\\
   0, & \text{otherwise}.
 \end{cases}
\]
This sign records how a change in the creator's label at time $\tau$
contributes to the difference between its labels at $\tau_w$ and
$\tau_v$.
\begin{lemma}[Signed lifetime coefficient]\label{lem:signedhistory}
Fix a root-block lineage $Q$ and a time $\tau$ during its growing
period. Over every execution prefix ending at a completed update,
\begin{equation}\label{eq:signedhistory}
 \left|\sum_{e\text{ internal to descendants of }Q}\sigma_e(\tau)\right|\le2.
\end{equation}
\end{lemma}
\begin{proof}
Original entry order turns this time cut into a fixed key cut, with
orientation given by the lineage's original side
(\cref{lem:historycuts}). At any time only one fragment straddles it.
Splits can replace that fragment by a smaller straddling descendant,
or separate the two groups permanently.

In the first contributing access, at most one short internal edge
crosses the cut, by adjacency. The surviving members of the fragment
are consecutive in the residual list. Their internal long crossings
form a suffix of the alternating cut-search-path crossings, so their
signed sum has magnitude at most one by \cref{lem:cutpath}(ii).
The two geometric parts together have magnitude at most two.
Complete flushing leaves at most one member in this fragment; no
descendant can subsequently straddle the cut. This bounds the whole
lifetime coefficient. There is no accounting update between the short
and long parts of the forest.
\end{proof}
All internal edges must be kept in this signed sum. Removing the
equal-label edges can destroy alternation. Time-local uniqueness of
a straddling fragment alone would also be insufficient without flushing.

\paragraph{Capped creator records.}\label{sec:records}
Let $c_t(w)$ be the immutable creator label stored when the current
membership of $w$ began. If that membership began at time $\tau$ in
a block created by $u$, then $c_t(w)=g_\tau(u)$, with entry labels
evaluated as in \cref{sec:barriers}. The creator $u$ can differ from
the current owner after a shrinking transfer. The root has no
membership. Define
\begin{equation}\label{eq:cappedrecord}
 \Gamma_t=\sum_{w\text{ currently in a block}}\min\{c_t(w),g_t(w)\},
 \qquad 0\le\Gamma_t\le(n-1)G.
\end{equation}
Only the potential is capped: the original stored number remains
unchanged by primary-label updates, splits, and shrinking transfers.
Let
\begin{equation}\label{eq:Tinternal}
 \Tint=\sum_{v\to w\text{ internal}}(g_t(v)-c_{t^-}(w)),
 \qquad J_0=n-1+2R,
\end{equation}
where $t$ denotes the frozen access snapshot. The number $J_0$ bounds
all initialization and pivot entries.

\begin{lemma}[Capped-record estimate]\label{lem:capped}
\begin{equation}\label{eq:cappedbound}
 \Tint\le GJ_0+W^++P^-.
\end{equation}
\end{lemma}
\begin{proof}
Separate record replacements, evaluated at frozen labels, from the
ensuing primary-label update. On an internal link $v\to w$,
$g_t(v)\le g_t(w)$, so the record replacement contributes
\[
 g_t(v)-\min\{c_{t^-}(w),g_t(w)\}
 \ge g_t(v)-c_{t^-}(w)
\]
to $\Gamma$. On a boundary link its contribution is at least
\[
 \min\{g_t(v),g_t(w)\}-g_t(w)
 =-[g_t(w)-g_t(v)]_+.
\]
The boundary-link record replacements therefore contribute at least $-W^+$.
Comparator events remove at most two old records per rotation, each
contributing at most $G$, so their record changes total at least
$-2RG$. Conversions, splits, and history-preserving transfers do not
change records. Finally, decreasing a primary label reduces
$\min\{c,g\}$ by no more than that decrease; all label updates
together contribute at least $-P^-$.

Telescoping these contributions gives
$\Gamma_f-\Gamma_0\ge\Tint-W^+-2RG-P^-$.
The left side is at most $(n-1)G$, proving the claim. This remains
valid if a winner also loses, because all records use the same
frozen labels and each losing membership is replaced only once.
\end{proof}

\paragraph{Counting all internal links.}
The next inequality collects the historical argument into one bound
whose right side is already controlled by the boundary estimates.
\begin{lemma}[Internal-link count]\label{lem:internalcount}
\begin{equation}\label{eq:internalcount}
 I\le5GJ_0+4(B+\Neq)+11V+4W.
\end{equation}
\end{lemma}
\begin{proof}
Fix an internal link $v\to w$ at snapshot $t$. Its old memberships
have entry times $\tau_v,\tau_w$ and the same original creator $u$.
Write
\[
 a_v=g_{\tau_v}(v),\qquad
 c_v=g_{\tau_v}(u),\qquad c_w=g_{\tau_w}(u).
\]
For access and pivot entries, the entry label is frozen before the
event's final primary-label update; initialization uses the completed
initial labels. Adding and subtracting these actual values gives
\begin{equation}\label{eq:unit}
 \begin{split}
 1={}&(1+c_v-a_v)+(a_v-g_t(v))\\
     &+(c_w-c_v)+(g_t(v)-c_w).
 \end{split}
\end{equation}
We sum this identity over all internal links and bound its four terms.

\emph{Entry terms.}
Group the first terms by the winner's old membership. Its entry
label $a$ and creator record $c$ remain fixed, so each internal win
contributes $1+c-a$. For a membership born at an internal link,
this contribution is nonpositive when the frozen endpoint labels
differ, and is one when they are equal. An initial or pivot
membership contributes at most $G$ per win. A membership born at a
boundary link $p\to z$ contributes at most
$1+[g(p)-g(z)]_+$ per win, using that birth's frozen labels.
Discard the nonpositive contributions from unequal-label internal
births. For the other birth types, apply the four-win bound of
\cref{lem:wins} to the displayed nonnegative upper bounds. This gives
\begin{equation}\label{eq:entrysum}
 \sum_{v\to w\text{ internal}}(1+c_v-a_v)
 \le4GJ_0+4(B+\Neq)+4W^-.
\end{equation}

\emph{Winner drift.}
The second sum is at most $4P^-$: each negative label change belongs
to at most one current membership and can be used by at most four
later internal wins of that membership. Label updates follow all
entries. A decrease at the end of an access therefore belongs to a
moving key's new membership; a survivor retains its existing one.

\emph{Creator differences.}
Let $s$ range over primary-label update times and write
$\Delta_s g(u)=g_{s^+}(u)-g_{s^-}(u)$.
Let $\mathcal G_u(s)$ contain the lineages with creator $u$ whose
growing periods contain $s$, and put
\[
 \kappa_{Q,s}
 =\sum_{e\text{ internal to descendants of }Q}\sigma_e(s).
\]
Expand each $c_w-c_v$ into the signed changes of the creator's label
between the two entry times and exchange the finite sums:
\[
 \sum_{v\to w\text{ internal}}(c_w-c_v)
 =\sum_u\sum_s\Delta_s g(u)
       \sum_{Q\in\mathcal G_u(s)}\kappa_{Q,s}.
\]
By \cref{prop:invariant}, $|\mathcal G_u(s)|\le3$, and
\cref{lem:signedhistory} gives $|\kappa_{Q,s}|\le2$. Consequently
\[
 \left|\sum_{v\to w\text{ internal}}(c_w-c_v)\right|
 \le6\sum_u\sum_s|\Delta_s g(u)|=6V.
\]
All internal links remain in these coefficients; their signs are
combined before absolute values are taken.

\emph{Record replacements.}
The fourth sum is exactly $\Tint$. Applying \cref{lem:capped}
and combining the four estimates yields
\[
 \begin{aligned}
 I
 &\le4GJ_0+4(B+\Neq)+4W^-+4P^-+6V+\Tint\\
 &\le5GJ_0+4(B+\Neq)+5P^-+6V+4W^-+W^+.
 \end{aligned}
\]
Since $P^-\le V$ and $4W^-+W^+\le4W$, this proves
\eqref{eq:internalcount}.
\end{proof}

\paragraph{Capped records in the four-key example.}\label{sec:workedrecords}
Initially the records of keys $1,2,3$ are $(1,1,1)$, so
$\Gamma=3$. Conversion changes their labels to $(1,3,2)$ and
leaves the records and $\Gamma$ unchanged. The whole access
uses these frozen values: $3\to2$ records $g(3)=2$ and $1\to3$
records $g(1)=1$. The final records are $(1,2,1)$ and final labels
$(1,2,2)$, giving $\Gamma=4$. Both links are internal, and
$\Tint=(2-1)+(1-1)=1$.

In the rotation of \cref{sec:workedrotation}, the record of key $1$
is replaced by a record of value $1$ for key $4$. Key $3$ keeps its
old record $1$ even though its label drops to $1$. Key $2$ keeps
record $2$ and label $2$. Thus the capped potential stays at $4$.
The potential uses current labels, while the records themselves
retain their historical values.

\subsection{Comparator-relative cost bound}\label{sec:closing}
The proof now has three independent inputs:
\[
 L=O(M),\qquad \Act\le4m+32\EE=O(M),\qquad
 V, B+\Neq, W=O(MG).
\]
The first is structural; the second counts nonempty shrinking blocks; the third
uses ordered cuts and label subtraction. None assumes a bound on $I$.

\begin{proof}[Proof of \cref{thm:comparison}]
The maintained history satisfies the single-membership, outward-entry,
and frozen-snapshot properties of \cref{prop:invariant}. Therefore
\cref{lem:internalcount} applies. Since $J_0=n-1+2R=O(M)$ and
$R\le\EE$, \cref{lem:boundarybudgets,lem:active,lem:births} give
\[
 I\le5GJ_0+4(B+\Neq)+11V+4W=O(MG).
\]
By exact assembly (\cref{lem:assembly}), the total touch count is
$C(X;h)\le3m+I+B=O(MG)$. All constants are absolute. For $n=1$,
the cost is exactly $m$, so the statement also holds.
\end{proof}

\begin{corollary}[Base competitiveness]\label{cor:base}
For every finite initial height profile on $[n]$,
\begin{equation}\label{eq:base}
 C(X;h)=O\bigl((n+\OPT_n(X))\log\log n\bigr).
\end{equation}
\end{corollary}
\begin{proof}
The normalized comparator of \cref{thm:normal} has
$H=O(\log(n+1))$, so $G=O(\log\log n)$. Its rotations have bounded depth, giving
$\EE=O(R)=O(\OPT_n(X))$. Also $m\le\OPT_n(X)$.
Apply \cref{thm:comparison}.
\end{proof}

The arbitrary-profile quantifier is essential. The structural argument
uses only comparisons of heights, and initialization creates $2n$
blocks and $n-1$ memberships regardless of the numerical profile.
Both initial potentials are bounded solely by $n,G$. The proof
therefore applies when interval contraction produces a height profile
that does not encode an initial BST. The next section uses this fact
to reduce the startup term to $O(n)$ and the competitive factor to
$O(\log\log k)$.

\section{Initialization and active-alphabet bounds}\label{sec:quotient}
The base comparison still multiplies $n$ by $G$. Unrequested keys explain why we can separate that startup term: extra touches inside an unrequested interval destroy old two-sided visibility records, and that loss happens only once. We retain one maximum-height representative per such interval rather than simply delete its influence on Greedy.

Let $X$ be nonempty, let $A=\{x_1,\ldots,x_m\}$, and put $k=|A|$. Partition $[n]$ into the active-key singletons and the nonempty maximal intervals containing no active key. Call the latter \emph{inactive gaps}, and let their number be $q\le k+1$. The number of partition intervals is
\[
 N=k+q\le2k+1.
\]
Let $\pi:[n]\to[N]$ be the order-preserving quotient map and set
$Y=(\pi(x_1),\ldots,\pi(x_m))$.
The partition may use the whole request sequence in the analysis; the actual algorithm and its initialization are unchanged. Every requested key is a singleton block of this partition. This condition is essential: maximum aggregation alone does not make arbitrary interval contractions commute with Greedy.

For example, if the only requested keys in $[10]$ are $3$ and $7$, the quotient groups are
\[
 [1,2],\quad\{3\},\quad[4,6],\quad\{7\},\quad[8,10].
\]
The three unrequested intervals remain present as maximum-height columns, while the two request columns are singletons. This partition uses knowledge of the sequence only to analyze an unchanged online execution.

\subsection{Exact quotient execution}
For a partition interval $J$, let $\bar h(J)=\max_{v\in J}h(v)$.
Choose the same new height in the original and quotient executions after
each request, greater than every current height in either execution.

\begin{lemma}[Exact inactive-gap quotient]\label{lem:quotient}
Run geometric Greedy on $X$ from $h$ and on $Y$ from $\bar h$. At every access, the quotient keys touched are exactly the partition intervals containing at least one key touched in the original execution. The quotient height of each interval remains the maximum height of its keys in the original execution.
\end{lemma}
\begin{proof}
Induct on accesses. The accessed interval is a singleton. For another interval $J$, let $\theta$ be the maximum height in the original execution between the requested key and the near boundary of $J$, excluding $J$ and including the requested key's old height. By induction this equals the maximum of the intervening quotient heights.

If $\bar h(J)\le\theta$, no key in $J$ can be a strict outward record. If $\bar h(J)>\theta$, the nearest key in $J$ attaining that maximum is a strict record relative to all preceding columns and is touched. These are precisely the quotient touch conditions. In a touched interval, at least one original coordinate and the quotient coordinate are raised to the same new maximum. An untouched interval changes neither. The invariant follows.
\end{proof}

For an inactive gap $J$, let $V_J(h)$ be the union of its strict
prefix-height and strict suffix-height record sets. To obtain this
union, scan $J$ from each end and include a key if its height exceeds
every height preceding it in at least one of the two scans. The first
key in each scan is always a record; visibility from either end
suffices for membership in $V_J(h)$. Define
\begin{equation}\label{eq:PhiA}
 \Phi_A(h)=\sum_{J\text{ inactive}}|V_J(h)|\le n-k.
\end{equation}

\begin{lemma}[Inactive-gap loss budget]\label{lem:quotientloss}
Let $U$ be the total number of touches on active keys in the original execution, including keys touched when not requested. Then
\begin{equation}\label{eq:quotientloss}
 C(X;h)\le\Phi_A(h)-\Phi_A(h_f)+2C(Y;\bar h)-U.
\end{equation}
In particular,
$C(X;h)\le(n-k)+2C(Y;\bar h)-m$.
\end{lemma}
\begin{proof}
We use the local outside-interval visibility argument
of~\cite{GPSW26}, stated here directly for finite height profiles.
Every request lies outside an inactive gap $J$. A key touched in $J$
was a strict height record when scanning $J$ from the end facing the
request, so it already belonged to $V_J$. An untouched key cannot
become a new prefix or suffix record: every old blocking height remains
at least as large, since heights only increase. Thus the new visible
set is a subset of the old one.

If $s_J>0$ keys in $J$ are touched, they are all raised to the same new
maximum. Every raised key strictly between the two extreme raised keys
has an equal-height blocker on each side, so it leaves $V_J$. Hence all
but at most two of the touched keys lose two-sided visibility. With
$\Delta_J=|V_J(h_{t-1})|-|V_J(h_t)|\ge0$, this gives
\[
 s_J\le2\ind{s_J>0}+\Delta_J.
\]
For $s_J=0$ the profile of $J$ does not change. Sum over times and gaps. By the exact quotient lemma, nonzero gap-touch events number $C(Y;\bar h)-U$. Adding the $U$ active touches and telescoping the losses gives \eqref{eq:quotientloss}; $U\ge m$ and \eqref{eq:PhiA} give the consequence. The argument uses only comparisons of heights and their common-maximum update, so no initial-tree assumption is needed.
\end{proof}
Only the local outside-interval visibility fact is imported. The exact quotient and the summation above are separate steps; no scaling recurrence or interleave bound is used.

\paragraph{Inactive-gap example.}\label{sec:workedquotient}
Take the eight-key initial spine from \cref{sec:workedcut}, now with
the entire request sequence consisting of one or more requests to $8$.
The only active key is $8$, and $J=\{1,\ldots,7\}$ is one inactive
gap. Its initial heights are $(0,-1,\ldots,-6)$, so every key in $J$
is a strict suffix record and $|V_J|=7$. The two quotient columns
have heights $(\bar h(J),h(8))=(0,-7)$.

On the first request, the original execution touches all eight keys and
the quotient touches both columns. Afterwards all heights equal $1$.
Only the two extreme keys $1,7$ remain in $V_J$, giving visibility
loss $7-2=5$. For this prefix, $U=1$ and the loss inequality is exact:
\[
  \underbrace{8}_{C(X;h)}
  =\underbrace{(7-2)}_{\Phi_A(h)-\Phi_A(h_f)}
   +2\cdot\underbrace{\vphantom{(7-2)}2}_{C(Y;\bar h)}
   -\underbrace{\vphantom{(7-2)}1}_{U}.
\]
Every later request touches only key $8$ in both executions. The tied
heights in $J$ block a new visit, and its five lost visibility records
cannot be charged again. The example exhibits precisely the one-time
startup work that the quotient separates from the competitive factor.

For the genuine initialization $h=-\depth_{T_0}$, with initial-tree height $H_0$, each one-sided strict record sequence has at most $H_0+1$ members. Hence
\begin{equation}\label{eq:startup}
 \Phi_A(-\depth_{T_0})\le\min\{n-k,\;2q(H_0+1)\}.
\end{equation}

\subsection{Projection of the optimum}
\begin{lemma}[Monotone column projection]\label{lem:projection}
If $P$ is a finite arboreally satisfied point set and $\pi$ is a nondecreasing map on its ordered key columns, then
\[
 \pi(P)=\{(\pi(x),t):(x,t)\in P\}
\]
is arboreally satisfied and has cardinality at most $|P|$.
\end{lemma}
\begin{proof}
Only pairs with distinct rows and columns need consideration. Let quotient corners have columns $i<j$ and rows $s\ne t$. Choose the rightmost point of $P$ on row $s$ mapping to $i$ and the leftmost point on row $t$ mapping to $j$. Their original columns are in increasing order.

A third point in their original closed rectangle exists by satisfaction. Its image lies in the quotient rectangle by monotonicity. It cannot coincide with the first quotient corner: its original point would then lie on row $s$, map to $i$, and be strictly to the right of the selected rightmost preimage. It cannot coincide with the second corner by the symmetric leftmost-preimage argument. Its image is consequently a third point satisfying the quotient pair. A map cannot increase cardinality.
\end{proof}
The extremal choice is essential; an arbitrary pair of preimages need not prevent a witness from collapsing onto a corner.

Apply \cref{lem:projection} to the positive-time touched set of an optimal original execution. Removing initialization rows preserves satisfaction among positive-time points, because their rectangles cannot contain negative-time witnesses. The projected set contains the quotient requests and has no more points. The geometric/BST correspondence, with free comparator initialization, gives
\begin{equation}\label{eq:OPTquotient}
 \OPT_N(Y)=O(\OPT_n(X)).
\end{equation}
An additive $O(N)$ setup term in a formulation of that correspondence is also absorbed: $N\le2k+1\le3m\le3\OPT_n(X)$. No particular initial comparator tree is asserted to survive the projection.

\subsection{Active-alphabet theorem and consequences}
The initial record count $\Phi_A$ and its bounds are now defined, so we can state the more precise refinement promised in the introduction.

\begin{theorem}[Active-alphabet refinement]\label{thm:alphabet}
There is an absolute constant $c_0$ such that, for every nonempty $X$ with $k$ distinct requested keys,
\begin{equation}\label{eq:alphabet}
\Greedy(X,T_0)
\le \Phi_A(-\depth_{T_0})+c_0\OPT_n(X)\llog(k).
\end{equation}
In particular, $\Phi_A(-\depth_{T_0})\le n-k$, so the cost is
$O(n+\OPT_n(X)\log\log k)$.
\end{theorem}

\begin{proof}[Proof of \cref{thm:alphabet}]
The quotient has $N\le2k+1$ keys. Apply \cref{cor:base} to its inherited maximum-height profile, which need not be an encoded initial BST. This gives
\[
 C(Y;\bar h)=O\bigl((N+\OPT_N(Y))\log\log N\bigr).
\]
Since $k\le m\le\OPT_N(Y)$ and $m\ge1$, we have
$N\le2m+1\le3\OPT_N(Y)$. Use \eqref{eq:OPTquotient} and $\log\log N=O(\log\log k)$ to obtain
\[
 C(Y;\bar h)=O\bigl(\OPT_n(X)\log\log k\bigr).
\]
Substitute into \eqref{eq:quotientloss}, discard the nonpositive terms $-\Phi_A(h_f)-U$, and absorb the factor two into an absolute constant. At $h=-\depth_{T_0}$ this is \eqref{eq:alphabet}. If $N=1$, the quotient cost is exactly $m$ and no normalization is needed.
\end{proof}

\begin{proof}[Proof of \cref{thm:mainalphabet,thm:main}]
For nonempty $X$, apply \cref{thm:alphabet} and use $\Phi_A(-\depth_{T_0})\le n-k\le n$ to obtain \eqref{eq:mainalphabet}. Since $k\le n$ and $\llog$ is increasing, this also proves \eqref{eq:main}. The empty sequence costs zero. If $m\ge n$, then $\OPT_n(X)\ge m\ge n$, so the startup term is absorbed and the ratio is $O(\log\log n)$ with the small-universe convention. The constant-factor BST realization of geometric Greedy, allowing a linear setup charge, preserves the same asymptotic form.
\end{proof}

\begin{corollary}[Profile-sensitive startup]\label{cor:startup}
If $H_0$ is the initial tree height and $q$ is the number of nonempty inactive gaps, then
\[
 \Greedy(X,T_0)
 \le\min\{n-k,2q(H_0+1)\}+c_0\OPT_n(X)\llog(k).
\]
For a balanced initial tree, the initialization term is $O(q\log(n+1))$, capped by $n-k$.
\end{corollary}
\begin{proof}
Combine \cref{thm:alphabet} with \eqref{eq:startup}.
\end{proof}

\begin{corollary}[Contiguous geometric windows]\label{cor:window}
Let $W$ be a nonempty contiguous window of a longer geometric Greedy execution, let $h_s$ be its actual starting profile, and let $A_W$ be its set of $k_W$ requested keys. Then
\[
 \operatorname{cost}_{\Greedy}(W)
 \le\Phi_{A_W}(h_s)+c_0\OPT_n(W)\llog(k_W).
\]
\end{corollary}
\begin{proof}
The proofs of \cref{lem:quotient,lem:quotientloss,lem:projection,thm:alphabet} apply to the finite inherited profile $h_s$, using \cref{cor:base} on the quotient. The original heights are not reset; only the analysis blocks and the free comparator for $W$ are initialized afresh.
\end{proof}
This statement is for contiguous windows of geometric touches, not arbitrary subsequences obtained by deleting intervening updates. It does not assert that a globally amortized physical simulation preserves every window estimate.

\section{Conclusion}\label{sec:conclusion}
We prove that online geometric Greedy is $O(\log\log n)$-competitive
by comparing its work directly with an evolving near-optimal BST.
The auxiliary heap expresses each access as one ordered forest of
short and long links. Persistent blocks make the representation stable
under comparator rotations, and a shrinking-block population count
bounds the number of block--access interactions. Ordered-cut estimates
control label variation and weighted boundary differences. A signed
entry identity, bounded historical reuse, and the capped-record
potential then account for all internal links,
giving a single factor equal to the logarithmic label range.

The inactive-gap quotient separates initialization from the competitive
factor and yields $O(n+\OPT_n(X)\log\log k)$ for $k$ distinct requested
keys. It also gives the profile-sensitive and contiguous-window bounds
of \cref{sec:quotient}. Whether Greedy is constant-competitive remains
open. A natural next question is whether a direct comparison can
reduce the dependence on the number of label levels.

\paragraph{Disclosure of AI use.}
ChatGPT was used to assist with (a) brainstorming about and simplifying
parts of the amortized analysis and (b) phrasing of the manuscript and
refining of the exposition. The author takes full responsibility for the correctness of the results.

\clearpage
\phantomsection

\end{document}